%% file: main.tex
\documentclass[sigconf]{acmart}
\AtBeginDocument{%
  }

\usepackage{algorithm}
\usepackage{algpseudocode}
\usepackage{wrapfig}
\usepackage{multirow}
\usepackage{amsmath}
\usepackage{adjustbox}
\usepackage{color, colortbl}
\usepackage{bbding}
\usepackage{subcaption}
\usepackage{makecell}
\usepackage{booktabs}
\usepackage{array}
\newcommand{\gc}{\cellcolor[gray]{0.915}}

\setcopyright{acmlicensed}

\copyrightyear{2026}
\acmYear{2026}
\setcopyright{cc}
\setcctype{by}
\acmConference[KDD '26]{Proceedings of the 32nd ACM SIGKDD Conference on Knowledge Discovery and Data Mining V.2}{August 09--13, 2026}{Jeju Island, Republic of Korea}
\acmBooktitle{Proceedings of the 32nd ACM SIGKDD Conference on Knowledge Discovery and Data Mining V.2 (KDD '26), August 09--13, 2026, Jeju Island, Republic of Korea}
\acmDOI{10.1145/3770855.3817760}
\acmISBN{979-8-4007-2259-2/2026/08}

\begin{document}

\title{Bypassing Copyright Protection in Diffusion-based Customization via Two-Stage Latent Feature Optimization}


\author{Ziang Xu}
\authornote{The first two authors contributed equally to this work.}
\orcid{0009-0002-7118-4794}
\email{220110723@stu.hit.edu.cn}
\affiliation{%
  \institution{Harbin Institute of Technology, Shenzhen}
  \city{Shenzhen}
  \state{Guangdong}
  \country{China}
}

\author{Wenbo Yu}
\authornotemark[1]
\orcid{0009-0004-8077-9487}
\email{wenbo.research@gmail.com}
\affiliation{%
  \institution{Tsinghua Shenzhen International Graduate School}
  \city{Shenzhen}
  \state{Guangdong}
  \country{China}
}

\author{Hongyao Yu}
\orcid{0009-0009-8525-1565}
\email{chrisqcwx@gmail.com}
\affiliation{%
  \institution{Tsinghua Shenzhen International Graduate School}
  \city{Shenzhen}
  \state{Guangdong}
  \country{China}
}

\author{Hao Fang}
\orcid{0009-0004-0271-6579}
\email{ffhibnese@gmail.com}
\affiliation{%
  \institution{Tsinghua Shenzhen International Graduate School}
  \city{Shenzhen}
  \state{Guangdong}
  \country{China}
}

\author{Jiawei Kong}
\orcid{0009-0001-4879-0668}
\email{kjw25@mails.tsinghua.edu.cn}
\affiliation{%
  \institution{Tsinghua Shenzhen International Graduate School}
  \city{Shenzhen}
  \state{Guangdong}
  \country{China}
}

\author{Bin Chen}
\authornote{Corresponding author.}
\orcid{0000-0002-4798-230X}
\email{chenbin2021@hit.edu.cn}
\affiliation{%
  \institution{Harbin Institute of Technology, Shenzhen}
  \city{Shenzhen}
  \state{Guangdong}
  \country{China}
}

\author{Hao Wu}
\orcid{0009-0000-9034-4330}
\email{wu-h22@mails.tsinghua.edu.cn}
\affiliation{%
  \institution{Tsinghua Shenzhen International Graduate School}
  \city{Shenzhen}
  \state{Guangdong}
  \country{China}
}

\author{Shu-Tao Xia}
\orcid{0000-0002-8639-982X}
\email{xiast@sz.tsinghua.edu.cn}
\affiliation{%
  \institution{Tsinghua Shenzhen International Graduate School}
  \city{Shenzhen}
  \state{Guangdong}
  \country{China}
}

\author{Zhiyong Wu}
\orcid{0000-0001-8533-0524}
\email{zywu@sz.tsinghua.edu.cn}
\affiliation{%
  \institution{Tsinghua Shenzhen International Graduate School}
  \city{Shenzhen}
  \state{Guangdong}
  \country{China}
}

\renewcommand{\shortauthors}{Ziang Xu et al.}

% 可选：设置页眉短作者名
% \renewcommand{\shortauthors}{Xu, Yu, Yu, Fang, Kong, Chen, Wu, Xia, and Wu}


\input{Sections/0_Abstract}




\keywords{Copyright Concerns, Adversarial Learning, Diffusion-based Customization}

\maketitle

\newcommand\kddavailabilityurl{https://doi.org/10.5281/zenodo.20508694}
\ifdefempty{\kddavailabilityurl}{}{
\begingroup\small\noindent\raggedright\textbf{Resource Availability:}\\
% please change the following context to include multiple artifacts if necessary, including data, models, code, etc.
The source code of this paper has been made publicly available at \url{\kddavailabilityurl}.
\endgroup
}

\input{Sections/1_Introduction}

\input{Sections/2_Related_Work}
\input{Sections/3_Method}
\input{Sections/4_Experiments}

\input{Sections/5_Conclusion}

% \begin{acks}
% \end{acks}

\begin{acks}
This work is supported in part by the National Natural Science Foundation of China under grant 62301189, 62576122,62571298, Guangdong Basic and Applied Basic Research Foundation under grant 2026A1515011139.
\end{acks}

\bibliographystyle{siam}
\bibliography{sample-base}

\appendix

\input{Sections/6_Appendix}

\end{document}

%% file: Sections/0_Abstract.tex
\begin{abstract}
    With the growing concerns over copyright infringement in diffusion-based customization, adversarial attacks have emerged as a prominent defense strategy to prevent malicious content forgery in personalized image generation. However, current defenses typically introduce persistent perturbations in the latent space of Latent Diffusion Models (LDMs), which remain susceptible to adaptive bypasses by adversaries. In this paper, we introduce Two-Stage Latent Feature Optimization (TS-LFO), an efficient and effective copyright-stealing attack against protected diffusion-based customization. We begin by observing that existing defenses primarily disrupt the mapping between input images and their latent representations, thereby degrading the model's ability to produce personalized outputs. To counteract this, TS-LFO restores the broken mapping through a two-stage optimization process. In the Latent Denoising Stage, we enhance semantic consistency between latent codes and input images by jointly minimizing a Latent-Image Alignment Loss and a Latent Diffusion Loss with timestep-dependent weights, effectively suppressing the high-frequency noise introduced by defenses. In the Latent Reconstruction Stage, we recover low-frequency semantic information using pixel-level constraints to refine the latent features. Extensive experiments show that TS-LFO consistently bypasses state-of-the-art (SOTA) copyright defenses and outperforms SOTA copyright attacks such as DiffPure, GrIDPure and IMPRESS across diverse settings.
\end{abstract}

%% file: Sections/1_Introduction.tex
\section{Introduction}

Recent years have witnessed the rapid development and widespread application of diffusion-based generative models \cite{croitoru2023diffusion, yang2023diffusion, cao2024survey, tan2024waterdiff}, with representative techniques such as DreamBooth \cite{ruiz2023dreambooth} and Textual Inversion \cite{gal2022image} pushing the boundaries of personalized visual content creation. These methods enable users to synthesize high-quality images conditioned on specific visual concepts by fine-tuning diffusion models on a small number of sample images, and the continuous optimization of diffusion model architectures and training strategies has further enhanced the quality and flexibility of personalized generation. However, the powerful generative capabilities of these models also bring about prominent security and intellectual property risks: malicious users can exploit diffusion-based customization tools to forge copyrighted visual content, counterfeit personal identities, and generate harmful information, which not only infringes on the legitimate rights and interests of content creators but also disrupts the healthy development of the generative AI ecosystem \cite{yu2024editable, fang2024privacy, qiu2024closer, fang2026enhancing}. In response to these risks, researchers have proposed adversarial perturbation-based copyright protection methods for Latent Diffusion Models (LDMs) \cite{liang2023adversarial, wang2024simac, liu2024disrupting}, the core architecture of most modern diffusion-based generative systems. The core principle of these protection methods is to add targeted adversarial perturbations to images based on the loss function of diffusion model personalized training. And through such perturbations, the model’s ability to perform unauthorized personalized generation for protected content is suppressed, thus enhancing copyright protection.

Although these copyright protection methods based on diffusion model personalized training loss have demonstrated preliminary effectiveness in countering naive unauthorized generation attempts, their underlying mechanisms have not been fully explored. Through in-depth research, we have made a key finding: the adversarial perturbations added by these copyright protection methods, while suppressing unauthorized personalized generation, will objectively disrupt the stable semantic mapping relationship between protected images and their latent space representations in LDMs (i.e., the latent-image mapping). This phenomenon constitutes the foundation for existing protection methods, creating a heavy reliance on it. It also implies potential vulnerabilities of existing copyright protection strategies, which our method aims to explore and utilize. And on this basis, we design a targeted attack method to bypass the current state-of-the-art copyright protection for diffusion-based personalized generation.

In this paper, we take this important finding as the core starting point, and propose a systematic and effective attack method named Two-Stage Latent Feature Optimization (TS-LFO) to bypass the adversarial perturbation-based copyright protection methods for LDMs. The TS-LFO framework is meticulously designed to exploit the vulnerability inherent in the objective phenomenon of latent-image mapping disruption, and it achieves the unauthorized reconstruction of protected content through two stages. In the Latent Denoising Stage, we focus on eliminating the high-frequency adversarial perturbations introduced by copyright protection methods and restoring the semantic consistency between perturbed latent features and original input images. To this end, we design a novel Latent-Image Alignment Loss and a Latent Diffusion Loss with timestep-varying weight factors, which align the denoising trajectory of perturbed latent features with the natural generation process of LDMs while effectively suppressing the interference of high-frequency perturbations. In the subsequent Latent Reconstruction Stage, we leverage the structural and low-frequency semantic information that is inevitably preserved in perturbed latent features (despite the disruption of latent-image mapping) and impose pixel-level constraints to restore the low-frequency semantic fidelity of protected content, ultimately realizing the accurate reconstruction of the original protected visual content from perturbed latent representations. In summary, our main contributions are as follows:

\begin{itemize}
    \item We propose a new copyright attack framework TS-LFO, a two-stage copyright attack that utilizes a Latent-Image Alignment Loss and a Latent Diffusion Loss with timestep-varying weight factors for high-frequency latent denoising, and utilizes pixel-level constraints for low-frequency semantic fidelity restoration.
    \item We analyze the robustness of copyright protections and the performance of copyright attack methods under diffusion-based customization methods. For copyright attack methods, we consider traditional adversarial defenses, advanced adversarial purification methods, and state-of-the-art (SOTA) copyright attack methods.
    \item Extensive experiments under various settings demonstrate that our TS-LFO can efficiently defeat SOTA copyright defenses for diffusion-based customization and outperform advanced copyright attack methods such as DiffPure, GrIDPure and IMPRESS.
\end{itemize}

%% file: Sections/2_Related_Work.tex
\section{Related Works}
% \subsection{Latent Diffusion}
% The Latent Diffusion Models (LDMs) \cite{rombach2022high} consist of an autoencoder \cite{zhang2018better} and a Unet \cite{ronneberger2015u} denoiser. For an input image $x$, the encoder $E(\cdot)$ first transforms $x$ into a latent feature $z=E(x)$. A decoder $D(\cdot)$ is trained to reconstruct $x$ from $z$, ensuring $D(E(x)) \approx x$. Subsequently, the Unet is trained. At each timestep $t$: \\
% 1. Sample $\epsilon \sim N(0,I)$. \\ 
% 2. The forward diffusion process generates $z_t$ by adding noise to $z$. \\
% 3. The pre-trained CLIP \cite{radford2021learning} Encoder encodes a text prompt $y$ into a conditional embedding $c$. \\
% 4. The Unet takes $z_t$, timestep t, and condition $c(y)$ as inputs to predict the noise $\epsilon$. \\
% 5. The training objective is to minimize the expected difference between the Unet’s prediction and the true noise $\epsilon$ across all timesteps: 

% \begin{equation}
%     L_{LDM}=\mathbb{E}_{z \sim E(x),y,\epsilon \sim N(0,I)}[ \Vert \epsilon-\epsilon_\theta(z_t,t,c(y)) \Vert ^2_2].
% \end{equation}

% After training, the model generates images via the reverse denoising process: \\
% 1. Sample $z_T \sim N(0,I)$. \\
% 2. Iteratively denoise $z_T$ over $T$ steps using the Unet to predict and remove noise, yielding $z_0$. \\
% 3. Decode $z_0$ into the final image $x=D(z_0)$ using the pre-trained decoder. \\

\subsection{Latent Diffusion Models (LDMs)}
Latent Diffusion Models (LDMs) \cite{rombach2022high, blattmann2023align, takagi2023high} typically comprises an autoencoder \cite{zhang2018better} and a Unet \cite{ronneberger2015u} denoiser. The autoencoder first encodes an input image $x$ into a latent representation $z=E(x)$ through its encoder $E(\cdot)$, while the decoder $D(\cdot)$ learns to reconstruct the original image such that $D(E(x)) \approx x$. Following this, the Unet denoiser undergoes training through a noise prediction paradigm. During each timestep $t$ in the training phase, a standard Gaussian noise sample $\epsilon \sim \mathcal{N}(0, I)$ is first generated. The latent representation $z$ then undergoes a forward diffusion process where the noise is progressively added to obtain $z_t$. Concurrently, the text prompt $y$ is encoded into a conditional embedding $c(y)$ using the pre-trained CLIP \cite{radford2021learning} encoder. These components (i.e., the noised latent $z_t$, the timestep information $t$, and the conditional embedding $c(y)$) are jointly fed into the Unet to estimate the added noise $\epsilon$. The model then optimizes its parameters $\theta$ by minimizing the expected squared error between the predicted noise and the true noise across all timesteps:

\begin{equation}
    \mathcal{L}_{LDM}=\mathbb{E}_{z \sim E(x),y,\epsilon \sim \mathcal{N}(0, I)}[ \Vert \epsilon-\epsilon_\theta(z_t,t,c(y)) \Vert ^2_2].
\end{equation}

During inference time, the image generation process initiates by sampling a Gaussian noise vector $z_T \sim \mathcal{N}(0, I)$. This noise vector then undergoes $T$ iterative refinement steps through the trained Unet, which progressively removes the estimated noise to recover the clean latent $z_0$. Finally, the pre-trained decoder reconstructs the generated image by mapping the denoised latent back to the pixel space, producing the output $x_{gen}=D(z_0)$.

LDMs have demonstrated remarkable capabilities in generative tasks \cite{fang2023gifd, chen2024editable, qiu2024mibench}, leading to their widespread adoption across multiple domains. In the field of image synthesis, LDMs have shown superior performance in generating high-quality and diverse samples \cite{liu2023more}. Their applications can also extend to video synthesis, where they enable temporally coherent generation through novel architectural designs \cite{esser2023structure}. For image editing tasks, LDMs provide flexible manipulation of visual content while preserving semantic consistency \cite{kawar2023imagic}. Notably, the success of LDMs has further expanded to 3D content creation, offering efficient text-to-3D generation pipelines \cite{poole2022dreamfusion}. These diverse applications highlight LDMs' versatility and effectiveness in handling various generative challenges.

\subsection{Diffusion-based Customization}
With the increasing demand for customization, numerous fine-tuning methods for diffusion models have emerged, such as DreamBooth \cite{ruiz2023dreambooth} and Textual Inversion \cite{galimage}. These approaches allow users to embed visual concepts into diffusion models using image modalities as input, thereby enabling the models to generate images highly similar to the input images. 

\textbf{Textual Inversion} embeds concepts by training a token in the prompt, learning the embedding corresponding to that token without modifying the weights of the diffusion model. 

In contrast to Textual Inversion, \textbf{DreamBooth} primarily embeds concepts by fine-tuning the weights in the Unet. First, DreamBooth introduces a prior loss to prevent overfitting. Second, it fine-tunes only the weights in the cross-attention layers of the diffusion model, as these layers are more critical for customization. Consequently, DreamBooth demonstrates superior performance and is currently the most popular customization method.

\subsection{Copyright Protection by Adversarial Samples against Diffusion-based Customization}

AdvDM \cite{liang2023adversarial} first proposed the idea of leveraging adversarial samples \cite{fang2025one, xiao2026diffusion, fang2024clip} for copyright protection against diffusion-based customization. For an input image $x$, the goal of AdvDM is to optimize a perturbation $\delta$ that satisfies:
\begin{equation}
    \|\delta\|_\infty < \epsilon_{LDM}, \quad \delta^* = \arg\max_\delta \mathcal{L}_{LDM}(x + \delta),
\end{equation}
which will result in the adversarial sample $x_{adv} = x + \delta^*$. When generating images $x_{gen}$ via diffusion-based customization methods (e.g., DreamBooth \cite{ruiz2023dreambooth}, Textual Inversion \cite{gal2022image}) using $x_{adv}$, the generated image $x_{gen}$ will exhibit significant discrepancies from the original image $x$, thereby achieving enhanced copyright protection.

Since then, many works \cite{wang2024simac, liu2024disrupting} have been proposed to enhance the defense robustness on the basis of AdvDM. SimAC \cite{wang2024simac} improves upon AdvDM by adding a Feature Interference Loss to the original loss function and employing a greedy timestep selection strategy, achieving better performance on facial images. DisDiff \cite{liu2024disrupting} incorporates a Cross-Attention Erasure module and a Merit Sampling Scheduler, which can also surpass AdvDM in terms of defense effects.

\textbf{Our Insights.} These methods fundamentally disrupt the mapping between the latent space and input images by misleading the autoencoder and the Unet \cite{yu2025gi}. Inspired by this, we propose TS-LFO, which bypasses existing copyright defenses through reconstructing the mapping between the latent space and input images, thereby enabling image copyright theft.

\subsection{Copyright Attacks Based on Purification}
\label{subsec:copyright_attacks_purification}
While copyright protection methods such as AdvDM have achieved notable success, numerous copyright attack techniques have emerged and can circumvent these protections, thereby facilitating the theft of original image copyrights. Notable examples include DiffPure \cite{nie2022diffusion}, GrIDPure \cite{zhao2024can}, and IMPRESS \cite{cao2023impress}. These attacks primarily leverage purification mechanisms to remove protective perturbations applied to images, restoring them to a state suitable for unauthorized use by malicious attackers.

\textbf{DiffPure} is an adversarial purification method that elegantly utilizes the noising and denoising stages of diffusion models to remove perturbations. During the noising stage, Gaussian noise is added to the adversarial example (i.e., the protected image). In the subsequent denoising stage, the pre-trained diffusion model's prior knowledge is leveraged to jointly remove both the added Gaussian noise and the adversarial perturbations, resulting in a purified, clean image. Although DiffPure was originally designed for adversarial purification, it has proven highly effective in stripping away perturbations introduced by copyright protection mechanisms, thus functioning as a potent copyright attack.

\textbf{GrIDPure} is a copyright attack method derived from DiffPure. It enhances the purification process by dividing the adversarial input image into multiple overlapping grids. Each grid segment is individually purified using DiffPure with a relatively small diffusion timestep. The purified grid segments are then merged, typically via averaging in overlapping regions, to reconstruct the complete purified image. This grid-based approach aims to improve the efficiency and local accuracy of the purification process compared to processing the entire image at once.

% \textbf{IMPRESS} is a copyright attack method specifically designed to counteract copyright protection schemes. It addresses a key observation: when a defensively perturbed image is processed through an autoencoder, the reconstructed image often differs significantly from the original. To mitigate this, IMPRESS introduces a \emph{consistency loss} to minimize the discrepancy between the autoencoder's reconstruction of the purified image and the original image. Additionally, it employs a \emph{similarity loss} to ensure that the final purified image remains perceptually similar to the original protected image, thereby preserving its visual quality while effectively removing the protective perturbations.

\textbf{IMPRESS} is a specialized attack method designed for copyright protection techniques based on autoencoder architectures, such as PhotoGuard \cite{salman2023raising}. This method is built upon a key observation: such protection mechanisms significantly disrupt the semantic connection between the images generated by the diffusion model and the original reference image. To address this issue, IMPRESS introduces a \emph{consistency loss}, aiming to minimize the semantic-level discrepancy between the generated image and the reference image. To circumvent the high computational cost associated with the iterative denoising process of standard diffusion models, IMPRESS employs an efficient computational framework: it bypasses the full diffusion inference chain and instead calculates the loss function using only the lightweight encoder and decoder modules from a latent diffusion model. This design substantially reduces the memory and time overhead of the optimization process. Furthermore, the method incorporates a \emph{similarity loss} to constrain the perceptual-level structural similarity between the purified image and the original protected image, thereby ensuring the natural appearance and visual fidelity of the output.

% In contrast, \textbf{TS-LFO} is a method that builds upon the foundational ideas of IMPRESS. While both techniques incorporate loss functions related to the autoencoder, a key distinction lies in their optimization domains: TS-LFO operates primarily within the latent space of the autoencoder, whereas IMPRESS typically performs optimization directly on the full image pixel space. Furthermore, TS-LFO incorporates additional loss terms that consider the Unet component often found in diffusion models, aiming for a more comprehensive purification. IMPRESS, in contrast, does not explicitly utilize such Unet specific losses. For a detailed explanation of the TS-LFO methodology, please refer to the \hyperref[sec:method]{Method} section.

Existing copyright attack methods (e.g., DiffPure, GrIDPure, and IMPRESS) mostly adopt a generalized purification paradigm, which attempts to \textit{blindly} eliminate perturbations without understanding or exploiting the specific underlying mechanisms of the protection methods. Such non-targeted design strategies inherently limit the upper bound of attack effectiveness. In contrast, a profound understanding of what the protection methods actually disrupt would enable a targeted rectification of the damaged content, thereby facilitating the design of a more efficient and potent attack framework.

To address the aforementioned limitations, this paper takes the two core sub-modules of diffusion-based customization fine-tuning as a starting point to systematically investigate the internal impact mechanisms of current copyright protection methods. Guided by these insights, we propose a highly targeted attack framework named Two-Stage Latent Feature Optimization (TS-LFO). 

%% file: Sections/3_Method.tex
\section{Method}
\label{sec:method}

%这里暂时将他加入到论文中，避免审稿人不懂版权攻击的场景
%或许需要修改这段的表述
% \subsection{Definition and Evaluation of Copyright Attack}
\subsection{Threat Model and Evaluation of Copyright Attack}

% \begin{algorithm}[b]
% \caption{Copyright Attack Evaluation}
% \small
% \label{alg:copyright_attack}
% \begin{algorithmic}[1] 
% \Require 
%     \State $x$: Original copyrighted image
%     \State $G(\cdot)$: Pre-trained diffusion-based customization generator
%     \State $\mathrm{Copyr\_Def}(\cdot)$: Defense method exerting adversarial perturbations on $x$
%     \State $\mathrm{Copyr\_Atk}(\cdot)$: Attack method altering the generation process of $G(\cdot)$
%     \State $M(\cdot,\cdot)$: Measurement function for customization performance metric (e.g., FID)
% \Ensure
%     \State $\mathrm{performance}$: Customization performance measurement result
% \Procedure{EvaluateAttack}{$x, G, \mathrm{Copyr\_Def},  \mathrm{Copyr\_Atk}, M$} 
%     \State // Generate adversarial example:
%     \State \quad $x_{adv} \gets \mathrm{Copyr\_Def}(x)$
    
%     \State // Steer the generator's generation process:
%     \State \quad $G^* \gets  \mathrm{Copyr\_Atk}(G)$
    
%     \State // Generate output image:
%     \State \quad $x_{gen} \gets G^*(x_{adv})$
    
%     \State // Calculate performance:
%     \State \quad $\mathrm{performance} \gets M(x_{gen}, x)$
    
%     \State \Return $\mathrm{performance}$
% \EndProcedure
% \end{algorithmic}
% \end{algorithm}

\begin{figure}[b]
    \centering
    \includegraphics[width=\linewidth]{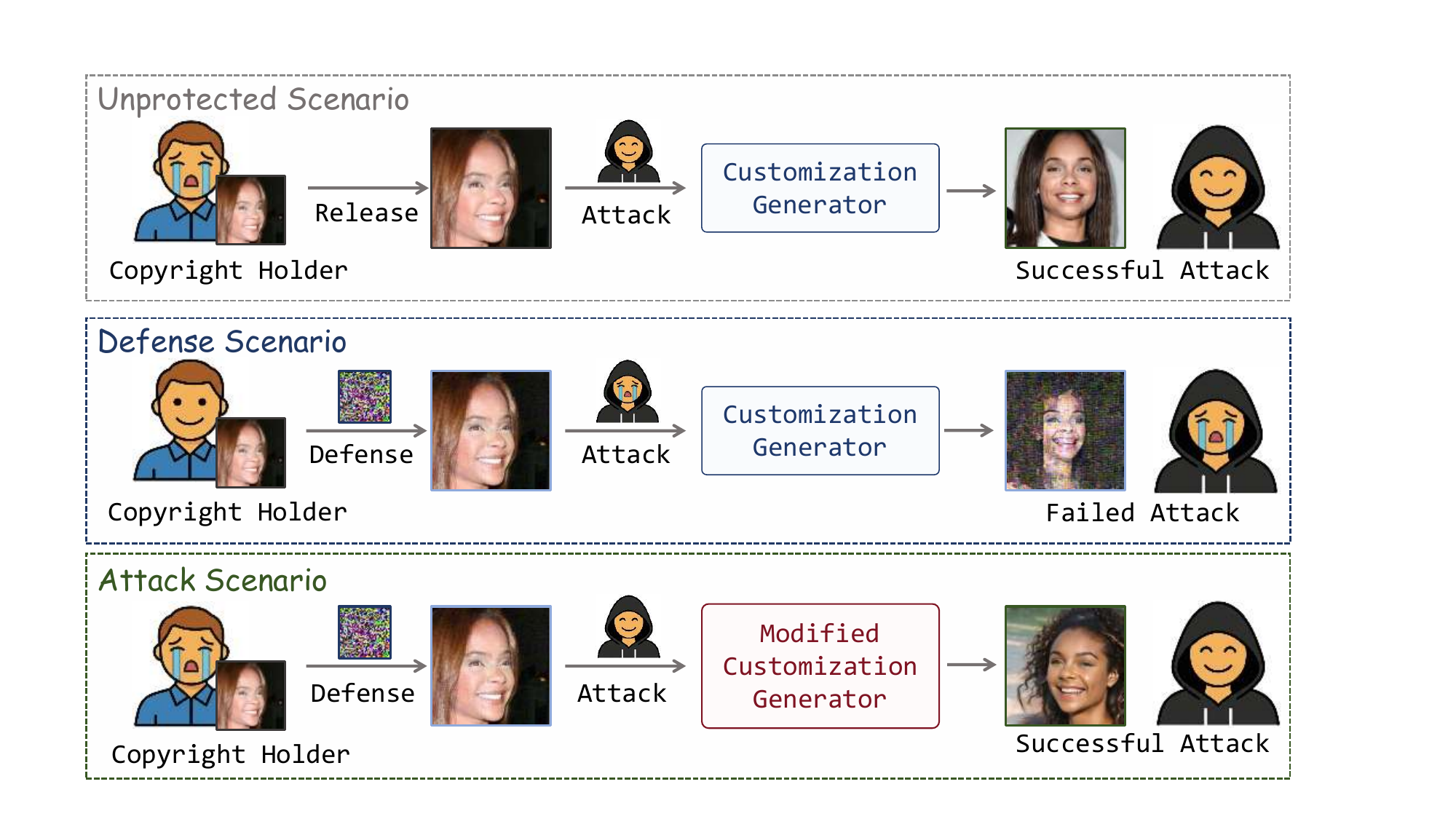}
    \caption{An illustration of a copyright attack and defence scenario. The three rows here represent the scenario without copyright protection, the scenario with only copyright protection, and the scenario where copyright protection and copyright attack act together. We use the similarity between the clean image $x$ held by the copyright holder and the image $x_{gen}$ finally generated by the attacker to measure the effectiveness of the copyright attack method.}
    \label{fig:intro}
\end{figure}

\begin{figure*}[t]
    \centering
    \includegraphics[width=\linewidth]{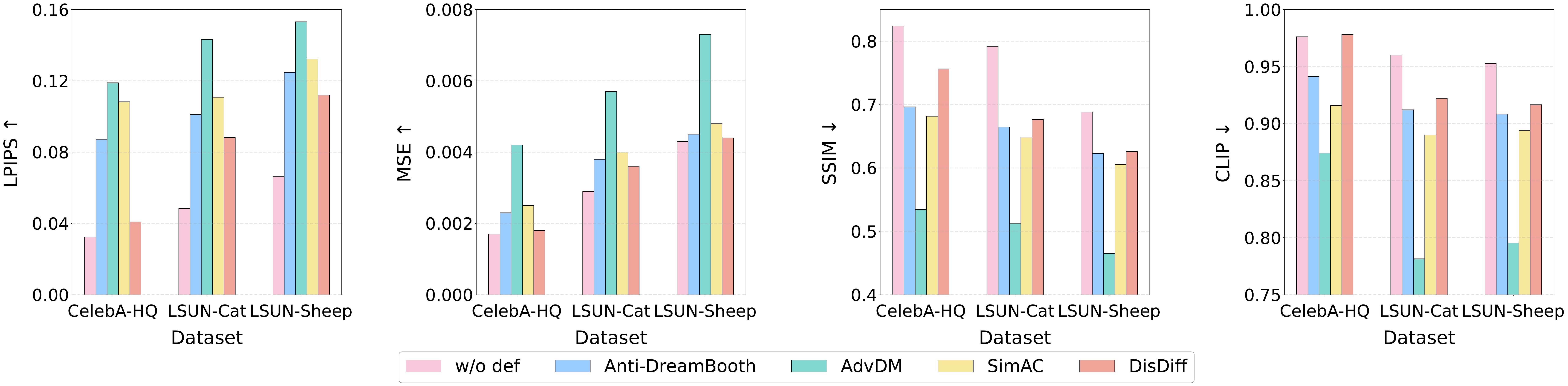}
    \caption{Comparison of the reconstructed image $x_{rec}=D(z)$ and its corresponding input image.}
    \label{tab:motivation1}
\end{figure*}

% Firstly, it is necessary to formally define the task of copyright infringement attacks. Suppose a copyright holder owns the rights to a clean image $x$ but wishes to prevent it from being learned by diffusion-based customization models. The holder therefore employs advanced copyright protection methods to generate a defensively perturbed image $x_{adv}$ from $x$, which is then publicly released. Then, a malicious attacker possessing a diffusion-based customization generator $G(\cdot)$ aims to generate images similar to $x$ but only has access to $x_{adv}$. The malicious attacker's objective is to appropriately steer the generation process to obtain $G^*(\cdot)$, so that the new generator $G^*(\cdot)$ can still produce the image $x_{gen}$ very similar to $x$ even when trained on the defensively perturbed $x_{adv}$.

% Thus, we define the above copyright attack process as a model modification mechanism $f \colon G(\cdot) \rightarrow G^*(\cdot)$, with its effectiveness measurable through the evaluation process depicted in Alg. \ref{alg:copyright_attack}.

% To address this defined problem, we propose the Two-Stage Latent Feature Optimization (TS-LFO) method. This method modifies the process of mapping input images $x_{adv}$ into the latent space within $G(\cdot)$ to obtain $x_{gen}$ with better similarity to $x$, thereby constituting a copyright attack. The rest of this section will elaborate on the implementation of each stage in TS-LFO.

Firstly, we formally define the task of copyright infringement attacks. Suppose a copyright holder owns a clean image $x$ but wishes to prevent it from being learned by diffusion-based customization models. To achieve this, the holder employs an advanced copyright protection method to generate a defensively perturbed image $x_{adv}$ from $x$ and publicly releases it. On this basis, we characterize the threat model from three distinct dimensions (i.e., attacker goal, attacker knowledge, and attacker capability) as follows:

\begin{itemize}
    \item \textbf{Attacker Goal:} To bypass the copyright protection mechanism and leverage the publicly available perturbed images for unauthorized personalized training of diffusion models, thereby generating content that exhibits high fidelity and similarity to the original clean image $x$.
    \item \textbf{Attacker Knowledge:} The attacker has full white-box access to the underlying architecture of the latent diffusion model (including the VAE and the Unet), but possesses no prior knowledge regarding the specific copyright protection methodology employed by the holder.
    \item \textbf{Attacker Capability:} The attacker can only access the publicly released defensive image $x_{adv}$ (with full operational control over it) while having no access to the original clean image $x$. Furthermore, the attacker retains the complete freedom to select arbitrary diffusion models to perform the infringing generation.
\end{itemize}

Under the aforementioned threat model, the objective of a malicious attacker possessing a diffusion-based customization generator $G(\cdot)$ is to appropriately steer the generation process to obtain $G^*(\cdot)$. This ensures that the new generator $G^*(\cdot)$ can still produce an image $x_{gen}$ highly similar to $x$, even when trained on the defensively perturbed image $x_{adv}$. The complete copyright attack and defence scenario is shown in Fig.~\ref{fig:intro}, with the last column representing the copyright attack scenario that this study focuses on.

Consequently, we define the above copyright attack process as a model modification mechanism $f \colon G(\cdot) \rightarrow G^*(\cdot)$, whose effectiveness can be quantitatively measured through comparing the similarity between $x$ and $x_{gen}$. Here, $x$ is the image owned by the copyright holder in the lower left of Fig.~\ref{fig:intro}, while $x_{gen}$ is the image generated by the attacker in the lower right of Fig.~\ref{fig:intro}.

To address this defined problem, we propose the Two-Stage Latent Feature Optimization (TS-LFO) method. This method modifies the process of mapping input images $x_{adv}$ into the latent space within $G(\cdot)$ to obtain $x_{gen}$ with better similarity to $x$, thereby constituting a copyright attack. The rest of this section will elaborate on the implementation of each stage in TS-LFO.

\subsection{Analysis on the Mechanism of Existing Copyright Protection Methods}
\label{subsec:analysis_defense}

To design more advanced and universal copyright attack method, the key lies in deeply understanding and deconstructing the operational mechanisms of existing copyright protection methods. In real-world scenarios, attackers can only access images perturbed by defense methods, denoted as $x_{adv}$. Therefore, this study aims to systematically analyze how $x_{adv}$ disrupts the functionality of core sub-modules during the personalized training of diffusion models (e.g., DreamBooth), thereby providing a theoretical foundation for designing targeted attack algorithms.

The personalized training of diffusion models primarily involves two core sub-modules: the VAE encoder \(E(\cdot)\) and the Unet denoiser $\epsilon_{\theta}(\cdot)$. The former is responsible for extracting compact latent features from input images, while the latter utilizes these features for denoising and concept learning. Based on this, we propose the following core research question: How do copyright protection perturbations separately disrupt the representational consistency of the encoder and interfere with the training dynamics of the denoiser? To address this question, we conduct empirical analyses from the following two perspectives.

\subsubsection{\textbf{Impact of Copyright Protection on the VAE Encoder}}
\label{subsubsec:impact_encoder}

We find that copyright protection disrupts the mapping between the protected image and its corresponding latent features in the latent space, preventing the encoder from extracting effective latent representations from the protected image $x_{adv}$. In other words, the feature $z = E(x_{adv})$ encoded from $x_{adv}$ is a ``corrupted'' representation, which cannot even accurately describe the visual content of $x_{adv}$ itself.

To quantitatively substantiate this finding, we design the following evaluation paradigm. For a given protected image $x_{adv}$, we reconstruct its latent feature $z$ back into an image $x_{rec} = D(z)$ via the VAE decoder \(D(\cdot)\). By measuring the discrepancy $M(x_{adv}, x_{rec})$ between the perturbed image \(x_{adv}\) and its reconstruction $x_{rec}$, we can directly assess the representational fidelity of the latent feature $z$. A larger discrepancy indicates that the perturbation causes more severe damage to the semantic consistency of the image itself and greater information loss during the encoding process. We employ a complementary set of metrics $M \in \{{LPIPS}, {MSE}, {SSIM}, {CLIP}\}$ for comprehensive evaluation, where an increase in LPIPS (perceptual difference) and MSE (mean squared error) signifies a larger discrepancy, while a decrease in SSIM (structural similarity) and CLIP (semantic similarity) scores also indicates a larger discrepancy.

We conducted experiments on three datasets (i.e., CelebA-HQ, LSUN-Cat, LSUN-Sheep) using four representative copyright protection methods (i.e., Anti-DreamBooth, AdvDM, SimAC, DisDiff) and an unprotected baseline (i.e., w/o def). The results are shown in Fig. \ref{tab:motivation1}. The data reveals that compared to unprotected clean images (i.e., w/o def), images $x_{adv}$ processed by all protection methods exhibit significant representational inconsistency: their LPIPS and MSE metrics increase sharply, while their SSIM and CLIP scores drop substantially. Taking the CelebA-HQ dataset as an example, the LPIPS for w/o def is 0.0324, whereas the AdvDM method raises it to 0.1189: an increase of over 3.6 times. Its CLIP score drops from 0.9761 to 0.8742. This indicates that copyright protection severely disrupts the image-to-latent mapping in the latent space, causing the latent feature $z$ generated by the encoder to become a ``distorted'' representation that fails to effectively convey the visual information of the original image.

\begin{figure}[b]
    \centering
    \includegraphics[width=\linewidth]{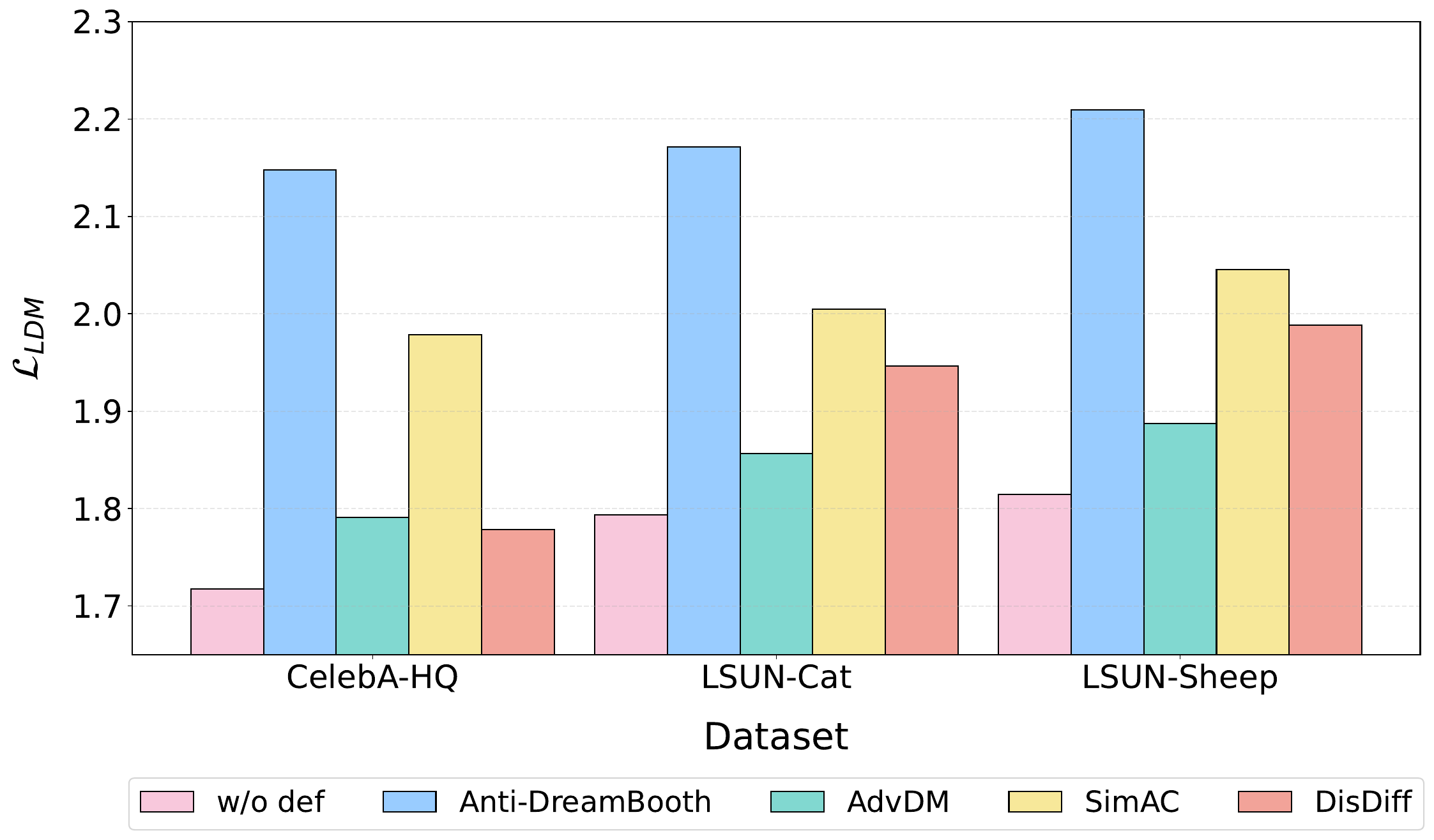}
    \caption{The impact of latent feature $z$ on the $\mathcal{L}_{LDM}$ under the influence of different copyright protections.}
    \label{tab:motivation2}
\end{figure}

\subsubsection{\textbf{Impact of Copyright Protection on the Unet Denoiser}}
\label{subsubsec:impact_unet}

Building upon the findings about the latent representations, we further investigate the impact of the corrupted latent feature $z$ on the subsequent training process of the Unet denoiser. In diffusion model training, the objective of the Unet is to predict the added noise based on a noisy latent feature. When the input training sample is the semantically inconsistent $z$, we hypothesize that it provides contradictory and unstable learning signals to the Unet, thereby interfering with the model's extraction of the core features of the target concept. This ultimately manifests as optimization difficulties in the training process, namely, higher training loss.

To validate this conjecture, under the same experimental setup, we monitored the denoising loss values of the Unet during DreamBooth personalized training when using data processed by different protection methods. The results are shown in Fig. \ref{tab:motivation2}. The data indicates that all copyright protection methods significantly increase the Unet's training loss. Taking the CelebA-HQ dataset as an example, the baseline loss is 1.7173, while the Anti-DreamBooth method increases it to 2.1476: a rise of approximately 25\%. This result clearly supports our inference: the defects in latent representation induced by copyright protection perturbations are directly transmitted to the model training phase, significantly increasing the Unet's training loss and hindering its efficient and accurate learning of the target visual concept.

\subsection{The First Stage of TS-LFO: Latent Denoising Stage}

\subsubsection{\textbf{Latent-Image Alignment Loss: Restore the Mapping between Protected Images and Their Latent Representations.}} 

% \textbf{Latent-Image Alignment Loss.} 
The analysis in Section \ref{subsubsec:impact_encoder} validates the necessity to obtain latent features $z$ that closely preserve the semantic content of input. To enforce this alignment, we leverage the decoder $D(\cdot)$ from the autoencoder with a similarity metric $M(\cdot, \cdot)$, formulating the loss function as:
\begin{equation}
\mathcal{L}_{align} = M(D(z), x_{adv}),
\end{equation}
where $D(\cdot)$ denotes the decoder, $z$ is the optimized latent feature initialized as $z=E(x_{adv})$, and $M(\cdot, \cdot)$ quantifies image similarity. While advanced metrics like LPIPS and SSIM are considered, counterintuitive empirical results in Section \ref{4.4} demonstrate that traditional MSE (i.e., $\ell_2$ distance) outperforms them in this context.

% \begin{table*}[t]
% \caption{The impact of latent feature $z$ on the ${L}_{LDM}$ under the influence of different copyright protections}
% \label{tab:motivation2}
% \centering

% \begin{tabular}{@{}l*{5}{c}@{}}
% \toprule
% \multirow{2}{*}{Dataset} & \multicolumn{5}{c}{Method} \\
% \cmidrule(lr){2-6}
% & w/o def & Anti-DreamBooth & AdvDM & SimAC & DisDiff \\
% \midrule
% CelebA-HQ & 1.7173 & 2.1476 & 1.7907 & 1.9786 & 1.7785 \\
% LSUN-Cat & 1.7933 & 2.1712 & 1.8564 & 2.0045 & 1.9465 \\
% LSUN-Sheep & 1.8147 & 2.2095 & 1.8873 & 2.0453 & 1.9883 \\
% \bottomrule
% \end{tabular}

% \end{table*}

% \begin{figure}[b]
%     \centering
%     \includegraphics[width=\linewidth]{Figures/pipeline-new.pdf}
%     \caption{An overview of the Latent Denoising Stage, which utilizes a Latent-Image Alignment Loss $\mathcal{L}_{align}$ and a Latent Diffusion Loss $\mathcal{L}_{DM}$ with timestep-varying weight factors $\lambda_t$ for high-frequency latent denoising.}
%     \label{fig:pipeline}
% \end{figure}

\begin{figure}[b]
    \centering
    \includegraphics[width=\linewidth]{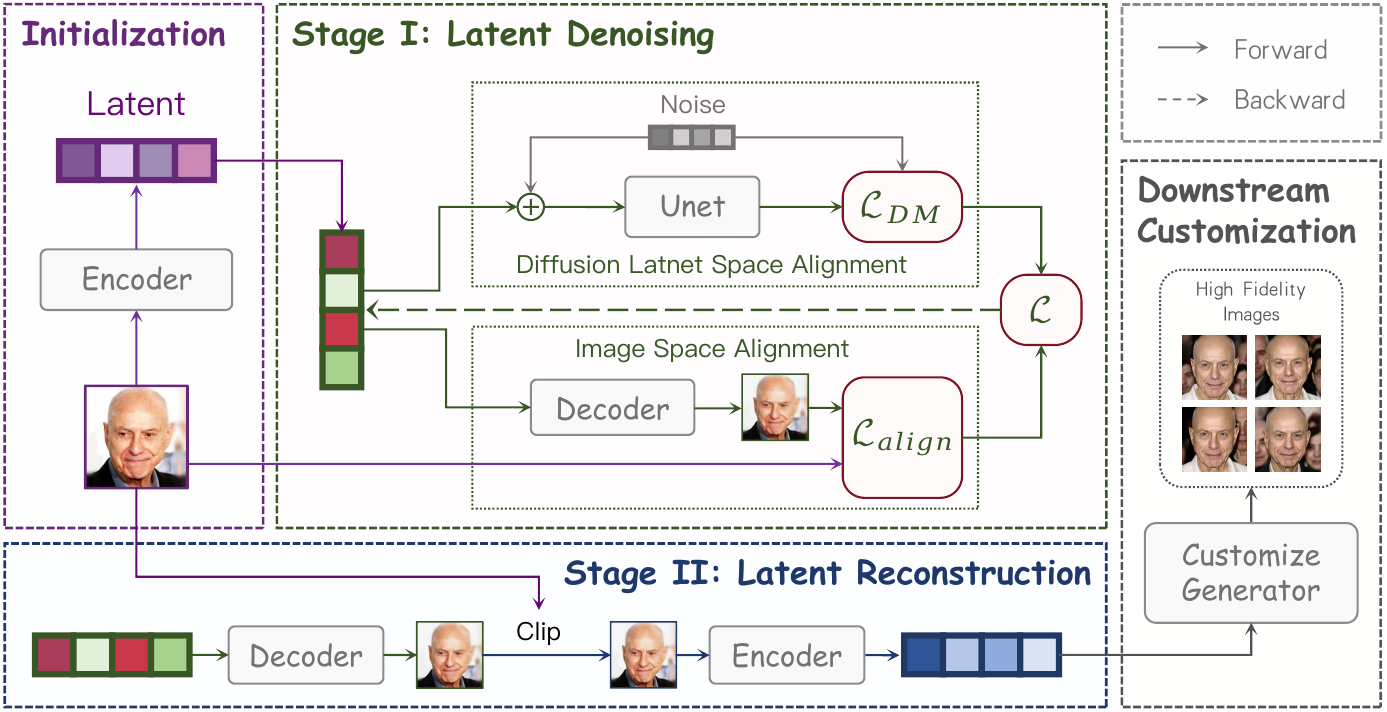}
    \caption{An overview of the complete TS-LFO pipeline, which consists of the Latent Denoising Stage, Latent Reconstruction Stage, and Downstream Customization Stage.}
    \label{fig:pipeline}
\end{figure}

% Crucially, the latent alignment loss universally counters state-of-the-art defenses including AdvDM, Simac, and DisDiff. These defenses operate by injecting perturbations through the LDM pipeline: 1) Input image $x$ undergoes adversarial perturbation, 2) LDM autoencoder $E$ encodes it into latent $z$ with amplified perturbations, 3) Diffusion model training on $z$ incurs destabilized loss. Our method directly attacks this vulnerability by optimizing $z$ to minimize $\mathcal{L}_{align}$, effectively neutralizing perturbation propagation in downstream training phases. 

% \textbf{Copyright protection perturbations disrupt Unet training dynamics.} 
\subsubsection{\textbf{Latent Diffusion Loss: Recover Unet Training Dynamics.}} 

% \textbf{Latent Diffusion Loss.} Current SOTA defenses like AdvDM \cite{liang2023adversarial}, SimAC \cite{wang2024simac}, and DisDiff \cite{liu2024disrupting} mainly come into effect by maximizing the training loss within the LDMs. To dismantle this defense strategy, we introduce a Latent Diffusion Loss to reduce the deviations of latent features:
% \begin{equation}
% \mathcal{L}_{DM} = \| \epsilon - \epsilon_\theta(z_t, t, c) \|_2^2,
% \end{equation}
% where $\epsilon \sim \mathcal{N}(0, I)$ denotes the random Gaussian noise added to the latent features, $\theta$ represents the LDMs' parameters, $t \in \{1, \ldots, T\}$ indicates the diffusion timestep, and $z_t$ is the noised latent feature at timestep $t$ computed through the forward process $z_t = \sqrt{\alpha_t} z + \sqrt{1 - \alpha_t} \epsilon$. The conditional vector $c$ is typically generated by the text encoders such as CLIP \cite{radford2021learning} and BERT \cite{kenton2019bert}, using prompts formatted as \texttt{"a photo of \{class\_name\}"}.

% This loss term strategically counteracts SOTA defense mechanisms by minimizing the denoising error $\mathcal{L}_{DM}$. As empirically validated in Section \ref{sec:main_experiments}, optimizing $\mathcal{L}_{DM}$ disrupts adversarial perturbations in the latent space, achieving significant perturbation elimination effects under various defenses.

% \textbf{Latent Diffusion Loss.} 

Through the quantitative analysis in Section \ref{subsubsec:impact_unet}, we confirm that the core negative impact of copyright protection perturbations is to significantly increase the denoising loss of the Unet during personalized training. This finding provides a direct optimization objective for the design of copyright attack methods: an effective purified image should serve as a ``high-quality training sample'' for the diffusion model, thereby reducing its training loss.

To this end, we propose the Latent Diffusion Loss, $\mathcal{L}_{DM}$, as the core supervisory signal to guide the optimization of the attack model. This loss directly simulates and minimizes a key loss term from the downstream DreamBooth training process.

Given a latent feature $z$ being trained, we simulate the forward noising step of the diffusion process in the latent space: randomly sample a timestep $t \sim \mathcal{U}(1, T)$, and add noise according to a predefined schedule to obtain the noisy latent feature $z_t$. Finally, we leverage a frozen, pre-trained Unet denoiser $\epsilon_{\theta}(\cdot)$ to predict the noise added to $z_t$ and compute the mean squared error against the actually added noise $\epsilon$. This is defined as the Latent Diffusion Loss:
\begin{equation}
\mathcal{L}_{DM} = \mathbb{E}_{t, \epsilon} \left[ \| \epsilon_{\theta}(z_t, t) - \epsilon \|_2^2 \right].
\end{equation}
The intuitive interpretation of this loss is: minimizing $\mathcal{L}_{DM}$ is equivalent to optimizing the latent feature \(z\) so that, from the perspective of the pre-trained diffusion model, $z$ itself is an easy-to-denoise, easy-to-learn feature. By directly reducing this proxy training loss, we force the optimized latent feature produced by our copyright attack method to fundamentally circumvent the interference that copyright protection imposes on the training dynamics, thereby laying the groundwork for subsequent successful personalized generation. This loss forms a closed logical loop with our previous experimental observations in Fig. \ref{tab:motivation2}: since copyright protection hinders learning by increasing the Unet training loss, directly optimizing against this loss constitutes the most effective attack pathway.

\subsubsection{\textbf{Timestep-Varying Weight Factors: Balancing the Effects of $\mathcal{L}_{align}$ and $\mathcal{L}_{DM}$.}} 

With the two loss terms $\mathcal{L}_{align}$ and $\mathcal{L}_{DM}$ defined, we formulate the total loss function with a weight factor $\lambda$:
\begin{equation}
\mathcal{L} = \mathcal{L}_{align} + \lambda \mathcal{L}_{DM}.
\end{equation}

In our preliminary experiments, we observe that for a fixed $\lambda$, $\mathcal{L}_{DM}$ naturally decays as timestep $t$ increases, causing imbalanced optimization: the training prioritizes $\mathcal{L}_{DM}$ reduction at smaller $t$ but switches focus to $\mathcal{L}_{align}$ at larger $t$. To enforce consistent optimization across all timesteps, we introduce timestep-dependent weight factors with linear scheduling:

\begin{equation}
\lambda_t = \lambda_l + \frac{(\lambda_r - \lambda_l)}{T} \cdot t,
\end{equation}
\begin{equation}
\mathcal{L} = \mathcal{L}_{align} + \lambda_t \mathcal{L}_{DM}.
\end{equation}

Here $t \sim \mathcal{U}(1, T)$ denotes the current timestep, $T$ is the total diffusion steps, and $\lambda_l$ and $\lambda_r$ represent the initial and final weight factors (typically set to $1\!\times\!10^{-4}$ and $3\!\times\!10^{-3}$ respectively). 

\subsection{The Second Stage of TS-LFO: Latent Reconstruction Stage}
\begin{figure}[b]
    \centering
    \includegraphics[width=\linewidth]{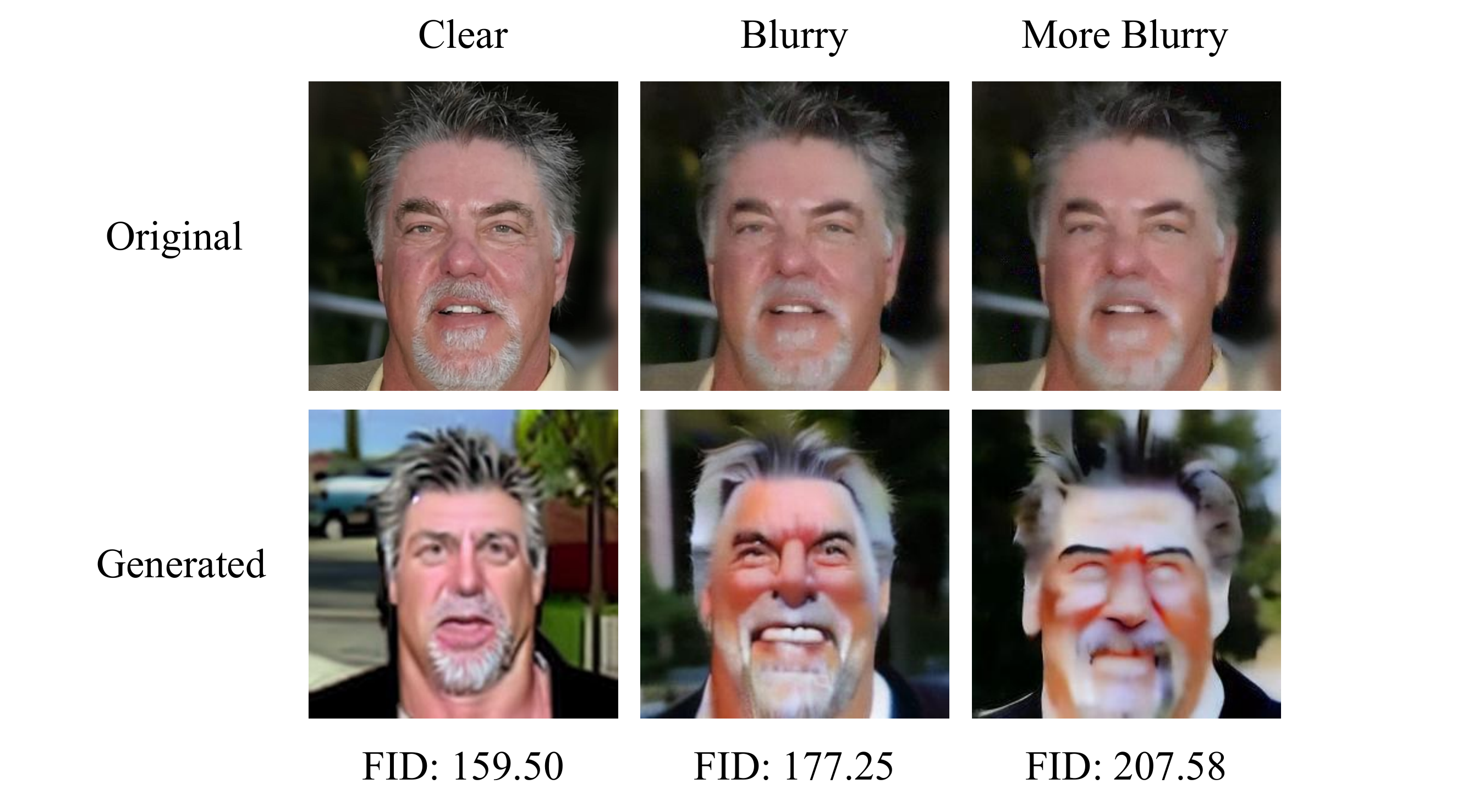}
    \caption{We train Textual Inversion with three image sets (i.e., clear images, blurry images, and more blurry images). This increasing trend in FID indicates that the quality of the input image determines the quality of the generated image.}
    \label{fig:quality}
\end{figure}

Unlike classification tasks, diffusion-based customization is a generative task where output quality critically depends on the input image quality. We validate this by training Textual Inversion with three image sets: clear images, blurry images, and more blurry images. As shown in Fig. \ref{fig:quality}, the FID scores between generated images and clear images exhibit a clear increasing trend from 159.50 to 177.25 and 207.58. This indicates that the input quality significantly affects the output quality.

After conducting the first stage, we obtain an optimized latent feature $z$ that retains semantic similarity to the input $x_{adv}$ but with reduced quality. To enhance $z$'s quality while preserving similarity to $x_{adv}$, we exert pixel-level constraints to obtain $z_{new}$:
\begin{equation}
\label{eq:9}
z_{new} = E(clamp(D(z), x_{adv}-\epsilon_{rec}, x_{adv}+\epsilon_{rec})),
\end{equation}
where $z_{new}$ is the reconstructed latent feature, $clamp(\cdot,\cdot,\cdot)$ is the cropped function, and $\epsilon_{rec}$ controls the reconstruction range. Here, $\epsilon_{rec}$ is a hyperparameter, typically set to $8/255$.

This constraint ensures the semantic meaning of optimized $z$ does not deviate significantly from $x$, thereby maintaining latent quality. Although minor defensive perturbations may persist in $z$, they have negligible impact on the LDMs' training. Experimental validation is provided in Section \ref{4.4}. An overview of the complete TS-LFO pipeline is illustrated in Fig. \ref{fig:pipeline}.

%% file: Sections/4_Experiments.tex
\section{Experiments}

\begin{table}[t]
    \caption{(DreamBooth) The performance of our method on SOTA defenses.}
    \label{tab:main_results_DB}
    \tiny
    \centering
    \tabcolsep=2pt
    \adjustbox{max width=\textwidth}{\begin{tabular}{ccccccccccc}
    \toprule
    \multirow{2}[0]{*}{Dataset} & \multirow{2}[0]{*}{Metric} & \multirow{2}[0]{*}{w/o def} & \multicolumn{2}{c}{Anti-DreamBooth} & \multicolumn{2}{c}{AdvDM} & \multicolumn{2}{c}{SimAC} & \multicolumn{2}{c}{DisDiff} \\
    \cmidrule(lr){4-5} \cmidrule(lr){6-7} \cmidrule(lr){8-9} \cmidrule(lr){10-11}
    & & & w/ def & \gc{\textbf{TS-LFO}} & w/ def & \gc{\textbf{TS-LFO}} & w/ def & \gc{\textbf{TS-LFO}} & w/ def & \gc{\textbf{TS-LFO}} \\
    \midrule
    \multirow{3}[0]{*}{LSUN-cat} & FID $\downarrow$ & 234.52 & 339.78 & \gc{\textbf{243.62}} & 401.08 & \gc{\textbf{249.94}} & 362.01 & \gc{\textbf{247.57}} & 327.16 & \gc{\textbf{244.12}} \\
    & CLIP$_{img}$ $\uparrow$ & 0.8228	& 0.7048 & \gc{\textbf{0.8135}} & 0.6353 & \gc{\textbf{0.7802}} & 0.6832 & \gc{\textbf{0.8093}} & 0.7313 & \gc{\textbf{0.8175}} \\
    & LPIPS $\downarrow$ & 0.4976 & 0.6278 & \gc{\textbf{0.5109}} & 0.5612 & \gc{\textbf{0.5134}} & 0.6237 & \gc{\textbf{0.5065}} & 0.5783 & \gc{\textbf{0.5072}} \\
    \cmidrule(lr){1-11}
    \multirow{3}[0]{*}{CelebA-HQ} & FID $\downarrow$ & 172.42 & 364.30 & \gc{\textbf{167.75}} & 237.00 & \gc{\textbf{181.34}} & 306.78 & \gc{\textbf{170.15}} & 175.03 & \gc{\textbf{165.86}} \\
    & CLIP$_{img}$ $\uparrow$ & 0.8020 & 0.6358 & \gc{\textbf{0.7519}} & 0.6878 & \gc{\textbf{0.7564}} & 0.6647 & \gc{\textbf{0.8031}} & 0.7886 & \gc{\textbf{0.8028}} \\
    & LPIPS $\downarrow$ & 0.4901 & 0.7000 & \gc{\textbf{0.5148}} & 0.5187 & \gc{\textbf{0.4687}} & 0.6333 & \gc{\textbf{0.4789}} & 0.5037 & \gc{\textbf{0.4757}} \\
    \cmidrule(lr){1-11}
    \multirow{3}[0]{*}{LSUN-sheep} & FID $\downarrow$ & 297.63 & 369.24 & \gc{\textbf{303.08}} & 417.33 & \gc{\textbf{306.16}} & 414.84 & \gc{\textbf{301.74}} & 395.20 & \gc{\textbf{304.83}} \\
    & CLIP$_{img}$ $\uparrow$ & 0.7942 & 0.7064 & \gc{\textbf{0.7700}} & 0.6244 & \gc{\textbf{0.7476}} & 0.6591 & \gc{\textbf{0.7827}} & 0.6883 & \gc{\textbf{0.7818}} \\
    & LPIPS $\downarrow$ & 0.4953 & 0.5961 & \gc{\textbf{0.5103}} & 0.5602 & \gc{\textbf{0.5179}} & 0.6020 & \gc{\textbf{0.5084}} & 0.5784 & \gc{\textbf{0.5052}} \\
    \bottomrule
    \end{tabular}}
\end{table}

\subsection{Experimental Setup}

\textbf{Datasets.} We choose three datasets: CelebA-HQ \cite{karras2017progressive} (face dataset), LSUN-cat \cite{yu2015lsun}, and LSUN-sheep \cite{yu2015lsun}.

\textbf{Defense Baseline.} We select open-source copyright protection methods using adversarial samples against diffusion-based customization: Anti-DreamBooth \cite{van2023anti}, AdvDM \cite{liang2023adversarial}, SimAC \cite{wang2024simac}, DisDiff \cite{liu2024disrupting}, and PhotoGuard \cite{salman2023raising}. 

\textbf{Attack Baseline.} We select SOTA adversial purification method DiffPure \cite{nie2022diffusion}, and the advanced copyright attack methods IMPRESS \cite{cao2023impress} and GrIDPure \cite{zhao2024can}. The method of adding Gaussian noise is also chosen as a representative of traditional copyright attacks.

\textbf{Metrics.} We evaluate similarity between generated and original images using FID \cite{heusel2017gans} (lower $\downarrow$ better), CLIP \cite{radford2021learning} Image-Image Similarity (higher $\uparrow$ better), and LPIPS \cite{zhang2018unreasonable} (lower $\downarrow$ better). Higher similarity indicates more successful copyright theft. For the face dataset, we additionally introduce two metrics: FDFR (lower $\downarrow$ better) and ISM (higher $\uparrow$ better). FDFR is designed to measure whether the generated results contain valid facial structures, obtained by calculating the proportion of images in which no face can be detected after processing by the RetinaFace \cite{deng2020retinaface} face detection model. ISM, on the other hand, is used to assess the identity consistency between the generated face and the target user. It is computed by employing the ArcFace \cite{deng2019arcface} model to extract identity feature vectors from the generated face and all original clean images of the user, followed by calculating the cosine similarity between the feature vector of the generated image and the average vector of the user's original feature set.

\textbf{Diffusion Models.} For DreamBooth, we use Stable Diffusion version 2.1 (SD-v2.1). For Textual Inversion, we use Latent Diffusion models. We use LDMs to compute $\mathcal{L}_{DM}$ in the first stage of TS-LFO.

\textbf{Implementation Details.} When optimizing latent features, we used the Adam optimizer with a learning rate set to 5e-3 and optimized for 3000 steps. For the CelebA-HQ dataset, we randomly selected 50 individuals, each of whom randomly selected 5 images. For the LSUN-cat and LSUN-sheep datasets, we randomly selected 50 individuals. For animal datasets lacking multiple images per individual, we augment each image through rotations at $-10^\circ$, $-5^\circ$, $0^\circ$, $5^\circ$, and $10^\circ$ to create 5 variants per identity. Next, we will use 5 images of the same individual to generate 50 images and compare the similarities and differences between the original clean images and the generated images, and then calculate their average. For all the methods mentioned above, we take their default parameters. For the DiffPure method, we set diffuse timestep as 100. For the adding Gaussian noise method, we set the mean to 0 and the standard deviation to 0.05. For all the defense methods mentioned above, we set the same noise budget of 16/255. All the experiments are conducted on NVIDIA GeForce RTX 3090 GPUs.

\begin{table}[t]
    \caption{(Textual Inversion) The performance of our method on SOTA defenses.}
    \label{tab:main_results_TI}
    \footnotesize
    \centering
    \tabcolsep=2pt
    \adjustbox{max width=\textwidth}{\begin{tabular}{ccccccccc}
    \toprule
    \multirow{2}[0]{*}{Dataset} & \multirow{2}[0]{*}{Metric} & \multirow{2}[0]{*}{w/o def} & \multicolumn{2}{c}{AdvDM} & \multicolumn{2}{c}{SimAC} & \multicolumn{2}{c}{DisDiff} \\
    \cmidrule(lr){4-5} \cmidrule(lr){6-7} \cmidrule(lr){8-9}
    & & & w/ def & \gc{\textbf{TS-LFO}} & w/ def & \gc{\textbf{TS-LFO}} & w/ def & \gc{\textbf{TS-LFO}} \\
    \midrule
    \multirow{3}[0]{*}{LSUN-cat} & FID $\downarrow$ & 256.75 & 404.62 & \gc{\textbf{284.33}} & 329.38 & \gc{\textbf{269.47}} & 303.13 & \gc{\textbf{267.34}} \\
    & CLIP$_{img}$ $\uparrow$ & 0.7902 & 0.6203 & \gc{\textbf{0.7661}} & 0.7123 & \gc{\textbf{0.7793}} & 0.7331 & \gc{\textbf{0.7779}} \\
    & LPIPS $\downarrow$ & 0.6157 & 0.6701 & \gc{\textbf{0.6275}} & 0.6483 & \gc{\textbf{0.6282}} & 0.6330 & \gc{\textbf{0.6327}} \\
    \cmidrule(lr){1-9}
    \multirow{3}[0]{*}{CelebA-HQ} & FID $\downarrow$ & 187.86 & 256.91 & \gc{\textbf{224.50}} & 286.98 & \gc{\textbf{204.80}} & 205.94 & \gc{\textbf{192.67}} \\
    & CLIP$_{img}$ $\uparrow$ & 0.7282 & 0.6212 & \gc{\textbf{0.6603}} & 0.5983 & \gc{\textbf{0.6807}} & 0.6769 & \gc{\textbf{0.6973}} \\
    & LPIPS $\downarrow$ & 0.5175 & 0.6770 & \gc{\textbf{0.6202}} & 0.6010 & \gc{\textbf{0.5455}} & 0.5578 & \gc{\textbf{0.5420}} \\
    \cmidrule(lr){1-9}
    \multirow{3}[0]{*}{LSUN-sheep} & FID $\downarrow$ & 271.45 & 414.45 & \gc{\textbf{294.34}} & 357.21 & \gc{\textbf{290.08}} & 328.04 & \gc{\textbf{291.39}} \\
    & CLIP$_{img}$ $\uparrow$ & 0.7494 & 0.5813 & \gc{\textbf{0.7385}} & 0.6656 & \gc{\textbf{0.7461}} & 0.6892 & \gc{\textbf{0.7482}} \\
    & LPIPS $\downarrow$ & 0.6061 & 0.6651 & \gc{\textbf{0.6233}} & 0.6556 & \gc{\textbf{0.6296}} & 0.6441 & \gc{\textbf{0.6266}} \\
    \bottomrule
    \end{tabular}}
\end{table}

\begin{table*}[t]
\caption{The performance of our method compared to the performance of SOTA copyright attacks on the CelebA-HQ dataset.}
\label{tab:main_results_face}
\centering
\setlength{\tabcolsep}{1.2pt}
\renewcommand{\arraystretch}{1.1}
\resizebox{\textwidth}{!}{%
\begin{tabular}{@{}l*{18}{c}@{}}
\toprule
\multirow{2}{*}{Method} & \multicolumn{3}{c}{w/o def} & \multicolumn{3}{c}{PhotoGuard} & \multicolumn{3}{c}{Anti-DreamBooth} & \multicolumn{3}{c}{AdvDM} & \multicolumn{3}{c}{SimAC} & \multicolumn{3}{c}{DisDiff} \\
\cmidrule(lr){2-4} \cmidrule(lr){5-7} \cmidrule(lr){8-10} \cmidrule(lr){11-13} \cmidrule(lr){14-16} \cmidrule(lr){17-19}
& FDFR & ISM & FID & FDFR & ISM & FID & FDFR & ISM & FID & FDFR & ISM & FID & FDFR & ISM & FID & FDFR & ISM & FID \\
\midrule
w/o atk & 0.1408 & 0.4650 & 172.4161 & 0.0360 & 0.3157 & 210.4718 & 0.1544 & 0.2563 & 364.3047 & 0.0836 & 0.2464 & 236.9997 & 0.2016 & 0.2843 & 306.7834 & 0.1636 & 0.4266 & 175.0311 \\
Noise & - & - & - & \textbf{0.0812} & 0.4365 & 178.0133 & 0.0828 & 0.4360 & 186.5673 & 0.0871 & 0.3435 & 206.6206 & 0.1076 & 0.4159 & 182.4842 & 0.1067 & 0.4566 & 176.6461 \\
IMPRESS & - & - & - & 0.0896 & 0.3272 & 189.2289 & \textbf{0.0684} & 0.3600 & 312.9439 & \textbf{0.0400} & 0.2831 & 235.0939 & 0.1276 & 0.3235 & 307.9065 & 0.1204 & 0.4406 & 184.1842 \\
DiffPure & - & - & - & 0.1024 & 0.4279 & 198.0258 & 0.1140 & 0.4281 & 202.8343 & 0.1580 & 0.3618 & 193.7353 & 0.1104 & 0.4202 & 205.4967 & 0.1528 & 0.4111 & 194.3153 \\
GrIDPure & - & - & - & 0.1672 & 0.4101 & 183.0769 & 0.1036 & 0.4462 & 201.7735 & 0.1736 & 0.3395 & 184.0426 & 0.1340 & 0.4457 & 201.4953 & 0.1476 & 0.4405 & 193.5929 \\
\gc{\textbf{TS-LFO}} & \gc{\textbf{0.1029}} & \gc{\textbf{0.4800}} & \gc{\textbf{169.3565}} & \gc{0.1012} & \gc{\textbf{0.4487}} & \gc{\textbf{171.6785}} & \gc{0.0988} & \gc{\textbf{0.4518}} & \gc{\textbf{167.7494}} & \gc{0.1084} & \gc{\textbf{0.3686}} & \gc{\textbf{181.3351}} & \gc{\textbf{0.1060}} & \gc{\textbf{0.4496}} & \gc{\textbf{170.1528}} & \gc{\textbf{0.1056}} & \gc{\textbf{0.4668}} & \gc{\textbf{165.8559}} \\
\bottomrule
\end{tabular}}
\end{table*}

\subsection{Main Experiments}
\label{sec:main_experiments}

This section validates the robustness of our proposed copyright attack against advanced defense mechanisms. We construct experiments on three datasets with distinct semantic differences: CelebA-HQ \cite{karras2017progressive} (face dataset), LSUN-cat \cite{yu2015lsun}, and LSUN-sheep \cite{yu2015lsun}, to verify our generalization across different image distributions. As shown in Table \ref{tab:main_results_DB} and Table \ref{tab:main_results_TI}, these defenses demonstrate significant effectiveness when not under attacks. Taking Textual Inversion as an example, we measure the average FID increase across three datasets (LSUN-cat, CelebA-HQ, LSUN-sheep) by subtracting the undefended baseline FID from the defended FID for each of the three SOTA defenses (AdvDM, SimAC, DisDiff). The resulting average increments are 119.97, 85.84, and 40.35, verifying the defenses' ability to disrupt identifiable features, consistent with existing literature.

Notably, applying our TS-LFO causes substantial degradation of defense effectiveness. Taking Textual Inversion as an example, post-attack FID values recover to 267.72, 254.78, and 250.47 respectively, showing reductions of 90.94, 69.74, and 28.57 compared to defended FID averages of 358.66, 324.52, and 279.04. These consistent improvements across all defenses demonstrate that our method effectively mitigates the protection of existing defensive mechanisms, restoring the attack's ability to generate identifiable images.

\subsection{Attack Experiments on Face Dataset}

To evaluate the efficacy of our TS-LFO method, we conducted experiments on the CelebA-HQ dataset, comparing against five copyright protection methods (i.e., PhotoGuard, Anti-DreamBooth, AdvDM, SimAC, and DisDiff) and four attack baselines (i.e., Gaussian Noise, IMPRESS, DiffPure, and GrIDPure). Results in Table~\ref{tab:main_results_face} demonstrate that TS-LFO achieves superior balance between attack effectiveness and generation quality.

On the core metrics measuring generation utility, our method exhibits strong attack performance based on the FDFR metric. While TS-LFO's FDFR values for defenses such as PhotoGuard, Anti-DreamBooth, and AdvDM are marginally higher than those of IMPRESS and Noise, they remain substantially lower than the original clean image baseline (0.1408). Moreover, under newer protections like SimAC and DisDiff, our approach significantly surpasses all baseline methods. Critically, even though TS-LFO was not optimized for autoencoder-based defenses (e.g., PhotoGuard), it achieves a notably lower FDFR (0.1012) than the unprotected clean image baseline, demonstrating that it not only removes perturbations but also actively enhances the latent representation to improve diffusion model compatibility.

On generation quality metrics, TS-LFO outperforms all baselines in both ISM and FID. For example, under PhotoGuard, TS-LFO attains an ISM of 0.4487 and FID of 171.6785, surpassing IMPRESS (ISM: 0.3272, FID: 189.2289). This highlights its ability to maintain high semantic fidelity while bypassing protections.

Furthermore, TS-LFO exhibits general image enhancement properties. When applied to clean images (w/o def), it improves all metrics: FDFR decreases to 0.1029, ISM increases to 0.4800, and FID reduces to 169.3565, confirming its role as an active representation optimizer rather than a mere perturbation filter.

Additionally, we observe that the Gaussian Noise method exhibits performance beyond expectations. Compared to other methods, Gaussian Noise not only shows certain advantages in the FDFR metric but also maintains decent performance in ISM and FID metrics. However, compared to our method, its FDFR does not show a significant advantage, and there remains a notable gap in ISM and FID metrics behind ours under the Anti-DreamBooth condition.

The attack results on more datasets are provided in Appendix~\ref{sec:attack_experiments}.

\subsection{Ablation Experiments}
\label{4.4}
This section validates the necessity of each core component through ablation experiments conducted on the CelebA-HQ dataset with AdvDM as the defense baseline using Textual Inversion.

\begin{table}
    \caption{Ablation Experiments on Timestep-Varying Weighting Coefficients.}
    \label{tab:ablation_lambda_t}
    \centering
    \adjustbox{max width=0.45\textwidth}{\begin{tabular}{@{}lccccc@{}}
        \toprule
        \tabcolsep=0.2pt
        $\lambda_t$ & 0 & $+\infty$ & $1\times10^{-4}$ & $3\times10^{-3}$ & \gc{\textbf{$1\times10^{-4}\sim3\times10^{-3}$}} \\
        \midrule
        FID $\downarrow$ & 241.73 & 260.59 & 237.20 & 231.55 & \gc{\textbf{224.50}} \\
        \bottomrule
    \end{tabular}}
\end{table}

\textbf{Timestep-Varying Weighting Coefficients.} To verify the dynamic balancing mechanism between the Latent-Image Alignment Loss $\mathcal{L}_{align}$ and the Latent Diffusion Loss $\mathcal{L}_{DM}$, we design five configurations. Through setting constant weight $\lambda_t=0$ to eliminate $\mathcal{L}_{DM}$ and $\lambda_t=+\infty$ to remove $\mathcal{L}_{align}$, we verify the necessity of both loss components. Additional experiments with static weights $1\times10^{-4}$ and $3\times10^{-3}$ demonstrate the need for dynamic adjustment. The proposed linear weight variation from $1\times10^{-4}$ to $3\times10^{-3}$ achieves the optimal FID score of 224.50 as shown in Table \ref{tab:ablation_lambda_t}.

\begin{table}
    \caption{Ablation Experiments on $\epsilon_{rec}$ Parameter in Latent Reconstruction Stage.}
    \label{tab:ablation_latent_reconstruction}
    \centering
    \begin{tabular}{@{}lcccc@{}}
        \toprule
        $\epsilon_{rec}$ & 4/255 & \gc{\textbf{8/255}} & 12/255 & 255/255 \\
        \midrule
        FID $\downarrow$ & 248.39 & \gc{\textbf{224.50}} & 227.77 & 230.16 \\
        \bottomrule
    \end{tabular}
\end{table}

\textbf{$\epsilon_{rec}$ Parameter in Latent Reconstruction.} To validate the balancing effect of $L_\infty$ constraint coefficient $\epsilon_{rec}$ in Equation \ref{eq:9}, we conduct experiments with four parameter settings. Disabling latent reconstruction by setting $\epsilon_{rec}=255/255$ confirms its necessity through a deteriorated FID value of 230.16. Comparative tests with $\epsilon_{rec}=4/255$ and $12/255$ reveal the trade-off between feature quality and perturbation magnitude, while the proposed $\epsilon_{rec}=8/255$ configuration achieves optimal balance with an FID score of 224.50 as detailed in Table \ref{tab:ablation_latent_reconstruction}.

\begin{table}
    \caption{Ablation Experiments on Metric Selection.}
    \label{tab:ablation_latent_image_alignment_loss}
    \centering
    \adjustbox{max width=0.5\textwidth}{\begin{tabular}{@{}lcccc@{}}
        \toprule
        Distance Metric & LPIPS & \gc{\textbf{MSE}} & CLIP$_{img}$ & SSIM \\
        \midrule
        FID $\downarrow$ & 235.94 & \gc{\textbf{224.50}} & 243.21 & 248.13 \\
        \bottomrule
    \end{tabular}}
\end{table}

\textbf{Metric Selection Analysis.} In Table \ref{tab:ablation_latent_image_alignment_loss}, our comparison of four similarity metrics for Latent-Image Alignment Loss $\mathcal{L}_{align}$ reveals unexpected findings. While the perception-level metrics (i.e., SSIM, LPIPS) and the semantic-level CLIP$_{img}$ demonstrate better similarity, traditional pixel-wise MSE achieves superior performance with an FID score of 224.50. This counterintuitive result stems from advanced metrics misinterpreting adversarial perturbations as legitimate features during optimization.

\begin{table}[htbp]
    \centering
    \caption{Ablation Experiments on Optimization Steps.}
    \label{tab:ablation_optimization_steps}
    \begin{tabular}{c c c c}
        \toprule
        Optimization Steps & FDFR $\downarrow$ & ISM $\uparrow$ & FID $\downarrow$ \\
        \midrule
        1000 & \textbf{0.0948} & 0.4373 & 196.1289 \\
        2000 & 0.1500 & 0.4307 & 169.7964 \\
        \gc{\textbf{3000 (default)}} & \gc{0.0988} & \gc{\textbf{0.4518}} & \gc{\textbf{167.7494}} \\
        4000 & 0.1296 & 0.4387 & 171.3990 \\
        5000 & 0.1104 & 0.4514 & 172.3843 \\
        \bottomrule
    \end{tabular}
\end{table}

\textbf{Optimization Steps Analysis.} We conduct an ablation study on optimization steps (1000$\sim$5000) to analyze its influence in Table \ref{tab:ablation_optimization_steps}. The experiments are conducted on the CelebA-HQ dataset using Dreambooth. With the steps rising from 1000 to 3000, FDFR first rises and then drops to the minimum (0.0988), ISM increases to the peak (0.4518), and FID decreases significantly to 167.7494, indicating better attack effectiveness, identity fidelity and generation quality. When the steps exceed 3000, FDFR rises, while ISM and FID fluctuate and degrade slightly, showing overfitting and unstable optimization. Overall, 3000 steps achieves the optimal trade-off between underfitting and overfitting, delivering the best comprehensive performance, which is set as the default configuration.

\subsection{Visualization Results and User Study}
Due to the length constraint policy, the visualization and user study results are presented in Appendix~\ref{sec:visual}.

% \subsection{Efficiency Experiments}
% \label{sec:efficiency_experiments}

% The time consumption for processing one image using our method, compared to other methods, is shown in the Table \ref{tab:efficency}. Comparatively, our method has the longest average processing time, which is a drawback. However, the effectiveness of our method surpasses all the methods mentioned above as shown in \ref{sec:attack_experiments}.

% \begin{table}
%     \caption{Time consumption for processing one image using our method, compared to other methods.}
%     \label{tab:efficency}
%     \centering
%     \adjustbox{max width=0.5\textwidth}{\begin{tabular}{@{}lcccccc@{}}
%         \toprule
%         Device & Gaussian noise & DiffPure & GrIDPure & IMPRESS & TS-LFO \\
%         \midrule
%         RTX 3090 & 0s & 8s & 64s & 163s & 653s \\
%         \bottomrule
%     \end{tabular}}
% \end{table}

%% file: Sections/5_Conclusion.tex
\section{Conclusion}
In this paper, we address the critical challenge of bypassing adversarial copyright defenses in diffusion-based customization by proposing the Two-Stage Latent Feature Optimization (TS-LFO) framework. Our investigation reveals that existing defense mechanisms rely on perturbing latent space representations to disrupt the alignment between input images and latent features to suppress personalized generation, leaving exploitable vulnerabilities. 

The TS-LFO framework systematically recovers the compromised latent-image mapping through a two-stage approach: the latent denoising stage employs adaptive loss functions to eliminate high-frequency perturbations while preserving semantic consistency, and the latent reconstruction stage restores low-frequency structural details via pixel-level constraints. Extensive experiments validate that our method successfully bypasses state-of-the-art defenses across diverse settings, achieving high-fidelity content generation even when models are protected by adversarial perturbations.

This work marks an effective solution for circumventing copyright defenses in diffusion-based customization, challenging the robustness of current adversarial protection strategies. Our findings highlight the need for more sophisticated defense mechanisms that account for feature vulnerabilities in latent spaces.

\textbf{Limitations, Future Direction, and Social Impacts.} Although the TS-LFO method demonstrates stronger copyright attack capabilities, its requirement of 6 minutes to process a single $512\times512$ image is indeed a drawback. However, compared to the DreamBooth training procedure (15 minutes) and the Textual Inversion training procedure (30 minutes), we argue that the time cost of the TS-LFO method is acceptable. Thus, TS-LFO remains a compelling choice for enhancing image generation quality. Moreover, TS-LFO bypasses protections including PhotoGuard, AdvDM, Anti-DreamBooth, SimAC, and DisDiff, alerting the research community. We find that protection researchers often overlook robustness evaluation against advanced attacks, focusing instead on traditional ones. We hope our work serves as a benchmark for evaluating protection robustness and anticipate more robust defenses against TS-LFO in the future. Ultimately, this study underscores the importance of balancing copyright protection with open innovation in the era of generative AI \cite{qiao2026wemath, zhang2024multitrust}, calling for collaborative efforts to establish ethical frameworks for responsible model customization.

%% file: Sections/6_Appendix.tex
\section{Algorithm of our method}

% Our proposed TS-LFO framework is outlined in Alg. \ref{alg:ts_lfo}. While the main manuscript assumes deterministic latent feature $z$ from autoencoder, practical implementations may involve probabilistic encoders like variational autoencoders (VAEs) that output Gaussian distribution parameters $(\mu, \sigma)$. To ensure compatibility, we further develop Alg. \ref{alg:ts_lfo_gaussian}, extending the optimization from deterministic $z$ to latent distribution parameters. This enhanced version maintains TS-LFO's core objective: \emph{reconstructing the mapping between input images and latent space while supporting diverse autoencoder architectures}.

Our TS-LFO framework is outlined in Fig. \ref{fig:pipeline}. While the main manuscript assumes deterministic latent feature $z$ from autoencoder, practical implementations may involve probabilistic encoders like variational autoencoders (VAEs) that output Gaussian distribution parameters $(\mu, \sigma)$. To ensure compatibility, we further develop Alg. \ref{alg:ts_lfo_gaussian}, extending the optimization from deterministic $z$ to latent distribution parameters. This enhanced version maintains TS-LFO's core objective: \emph{reconstructing the mapping between input images and latent space while supporting diverse autoencoder architectures}.

% Our proposed TS-LFO framework, which supports diverse autoencoder architectures, is fully presented in Alg. \ref{alg:ts_lfo_gaussian}. Considering that practical implementations of autoencoders may involve probabilistic encoders such as variational autoencoders (VAEs) that output Gaussian distribution parameters ($\mu$, $\sigma$), Alg. \ref{alg:ts_lfo_gaussian} is specifically designed to accommodate this scenario. It extends the optimization process from deterministic latent features $z$ (commonly assumed in autoencoder designs) to latent distribution parameters, while strictly maintaining the core objective of TS-LFO: reconstructing the mapping between input images and latent space.

\section{More Experimental Results}

\begin{table}[b]
    \caption{(DreamBooth) The FID performance of our method compared to the performance of SOTA copyright attacks on more datasets.}
    \label{tab:more_comparisons}
    \centering
    \tiny
    \tabcolsep=2pt
    \adjustbox{max width=\textwidth}{\begin{tabular}{ccccccccc}
    \toprule
    \multirow{2}[0]{*}{Defense} & \multirow{2}[0]{*}{Dataset} & \multirow{2}[0]{*}{w/o def} & \multirow{2}[0]{*}{w/ def} & \multicolumn{5}{c}{Attack} \\
    \cmidrule(lr){5-9}
    & & & & Noise & IMPRESS & DiffPure & GrIDPure & \gc{\textbf{TS-LFO}} \\
    \midrule
    \multirow{2}[0]{*}{Anti-DreamBooth} & LSUN-cat & 234.52 & 339.78 &	267.39 & 333.06	& 249.28 & 274.79 &	\gc{\textbf{243.62}} \\
    & CelebA-HQ & 172.42 & 364.30 & 186.57 & 312.94 &	202.83 & 201.77 &	\gc{\textbf{167.75}} \\
    & LSUN-sheep & 297.63 & 369.24 & 322.08 & 366.46 & 307.01 & 335.40 & \gc{\textbf{303.08}} \\
    \cmidrule(lr){1-9}
    \multirow{2}[0]{*}{AdvDM} & LSUN-cat & 234.52 & 401.08 & 334.16 & 398.40 & 255.95 & 319.12 & \gc{\textbf{249.94}} \\
    & CelebA-HQ & 172.42 & 237.00 & 206.62 & 235.09 & 193.74 & 184.04 & \gc{\textbf{181.34}} \\
    & LSUN-sheep & 297.63 & 417.33 & 385.16 & 417.01 & 311.20 & 355.94 & \gc{\textbf{306.16}} \\
    \cmidrule(lr){1-9}
    \multirow{2}[0]{*}{SimAC} & LSUN-cat & 234.52 & 362.01 & 274.19 & 364.37 & 251.37 & 280.06 & \gc{\textbf{247.57}} \\
    & CelebA-HQ & 172.42 & 306.78 & 182.48 & 307.91 & 205.50 & 201.50 & \gc{\textbf{170.15}} \\
    & LSUN-sheep & 297.63 & 414.84 & 331.21 & 419.27 & 305.03 & 328.96 & \gc{\textbf{301.74}} \\
    \cmidrule(lr){1-9}
    \multirow{2}[0]{*}{DisDiff} & LSUN-cat & 234.52 & 327.16 & 261.42 & 330.64 & 248.47 & 287.11 & \gc{\textbf{244.12}} \\
    & CelebA-HQ & 172.42 & 175.03 & 176.65 & 184.18 & 194.32 & 193.59 & \gc{\textbf{165.86}} \\
    & LSUN-sheep & 297.63 & 395.20 & 325.26 & 387.59 & 308.40 & 337.86 & \gc{\textbf{304.83}} \\
    \bottomrule
    \end{tabular}}
\end{table}

\begin{figure*}
    \begin{subfigure}{1\textwidth}
    \quad
        \begin{minipage}[t]{0.135\linewidth}
        \centering
        \includegraphics[width=2.44cm]{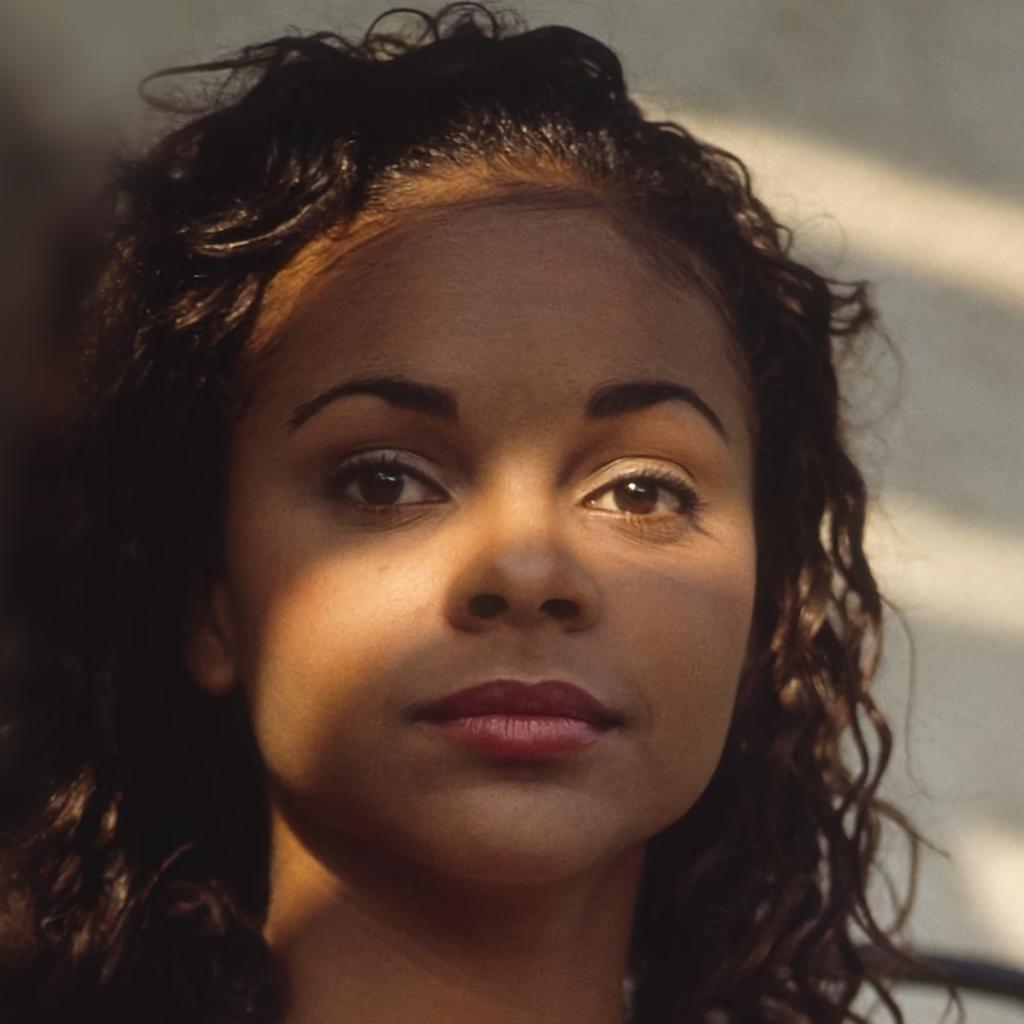}
        \centering
        \end{minipage}
        \begin{minipage}[t]{0.135\linewidth}
        \centering
        \includegraphics[width=2.44cm]{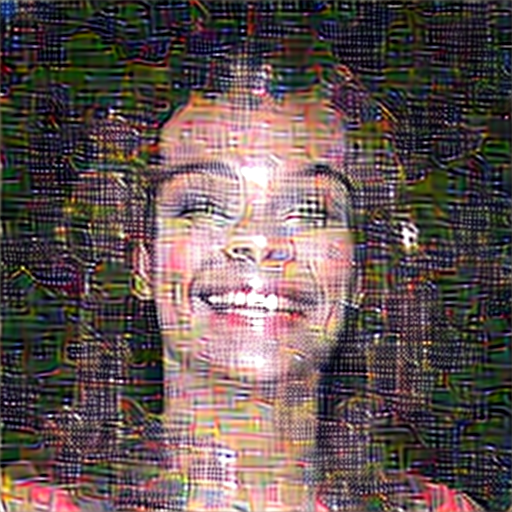}
        \centering
        \end{minipage}
        \begin{minipage}[t]{0.135\linewidth}
        \centering
        \includegraphics[width=2.44cm]{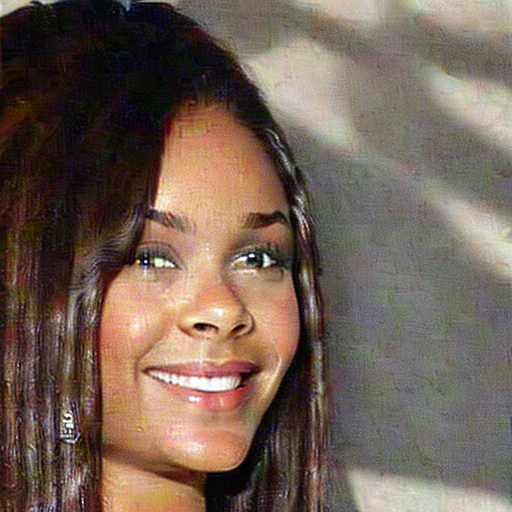}
        \centering
        \end{minipage}
        \begin{minipage}[t]{0.135\linewidth}
        \centering
        \includegraphics[width=2.44cm]{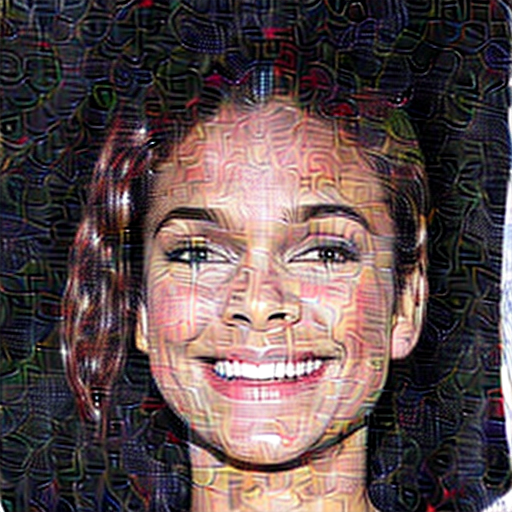}
        \centering
        \end{minipage}
        \begin{minipage}[t]{0.135\linewidth}
        \centering
        \includegraphics[width=2.44cm]{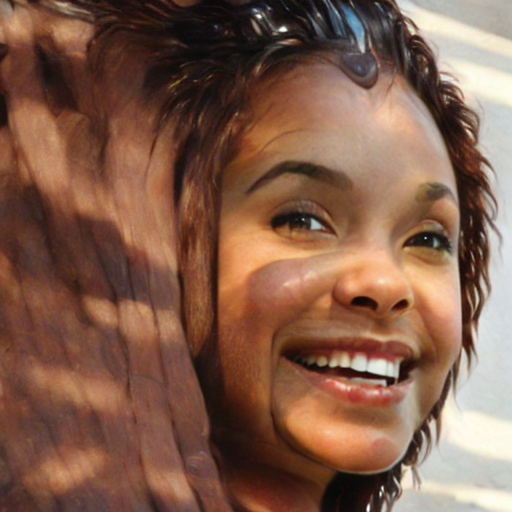}
        \centering
        \end{minipage}
        \begin{minipage}[t]{0.135\linewidth}
        \includegraphics[width=2.44cm]{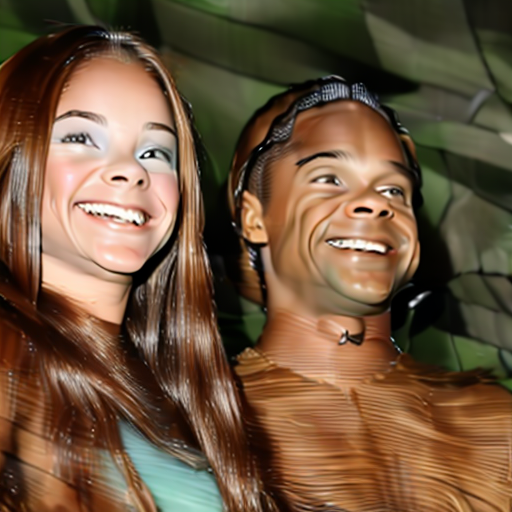}
        \centering
        \end{minipage}
        \begin{minipage}[t]{0.135\linewidth}
        \includegraphics[width=2.44cm]{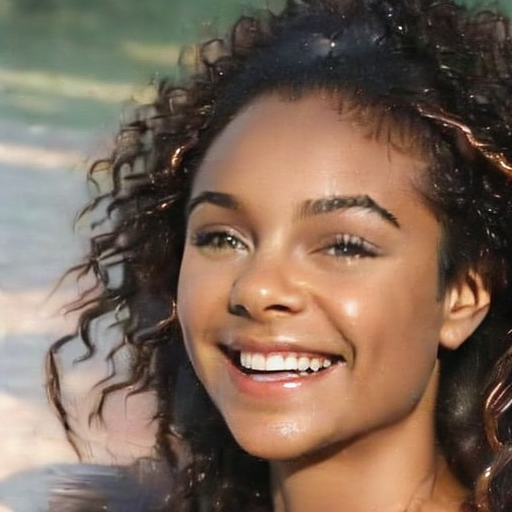}
        \centering
        \end{minipage}
        \end{subfigure}

    \begin{subfigure}{1\textwidth}
    \quad
        \begin{minipage}[t]{0.135\linewidth}
        \centering
        \includegraphics[width=2.44cm]{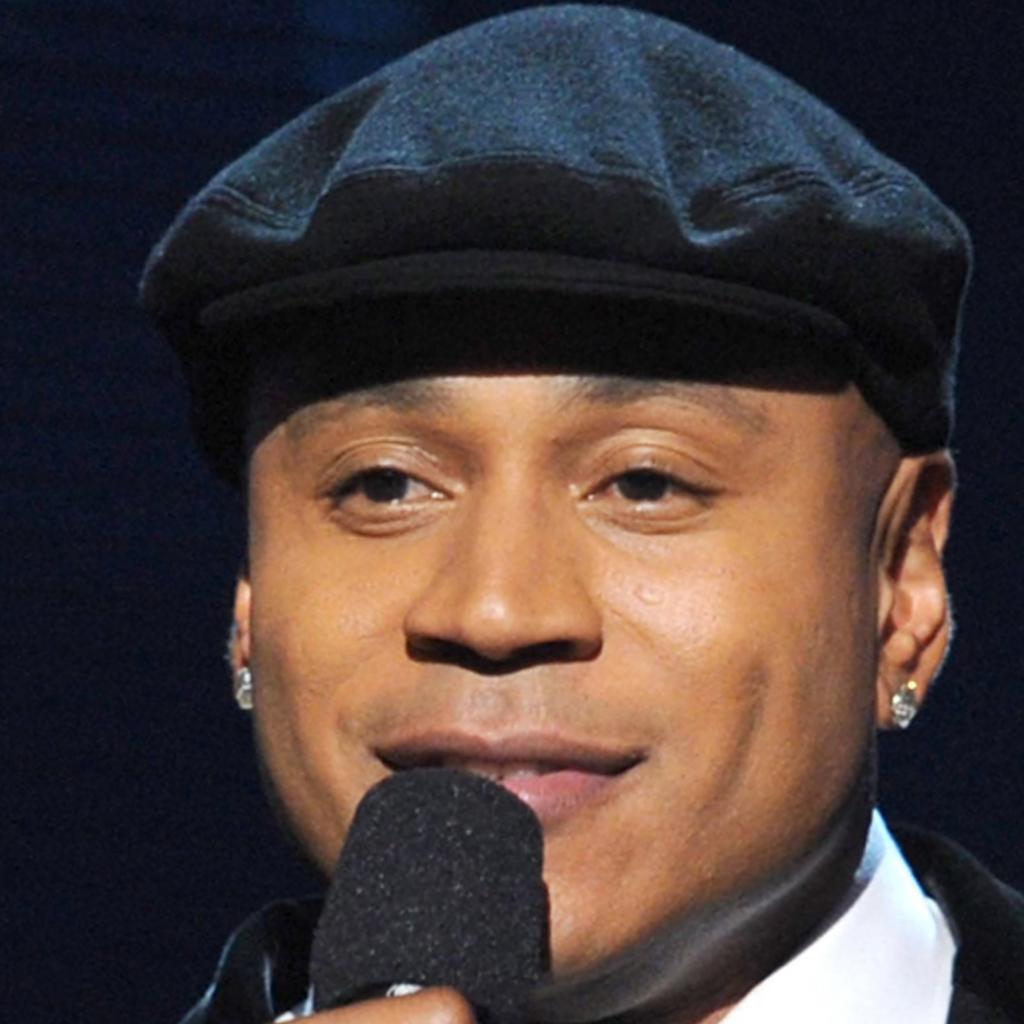}
        \centering
        \end{minipage}
        \begin{minipage}[t]{0.135\linewidth}
        \centering
        \includegraphics[width=2.44cm]{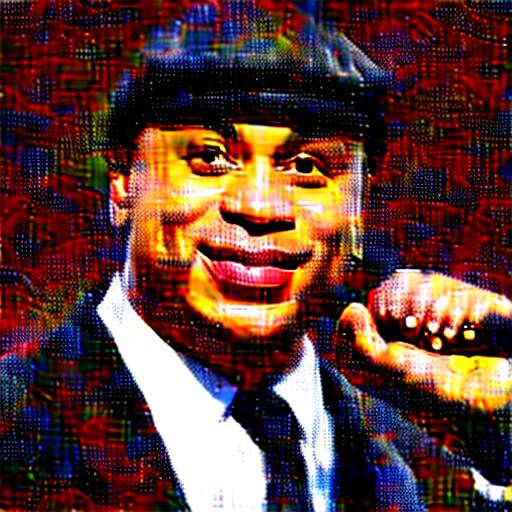}
        \centering
        \end{minipage}
        \begin{minipage}[t]{0.135\linewidth}
        \centering
        \includegraphics[width=2.44cm]{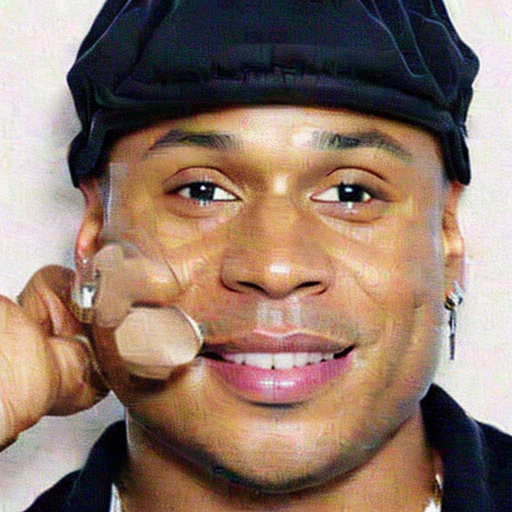}
        \centering
        \end{minipage}
        \begin{minipage}[t]{0.135\linewidth}
        \centering
        \includegraphics[width=2.44cm]{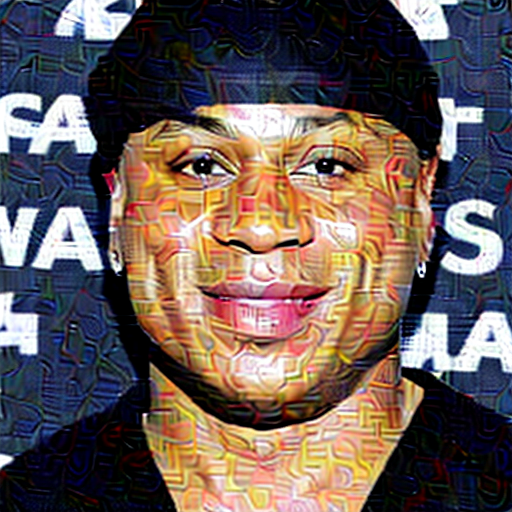}
        \centering
        \end{minipage}
        \begin{minipage}[t]{0.135\linewidth}
        \centering
        \includegraphics[width=2.44cm]{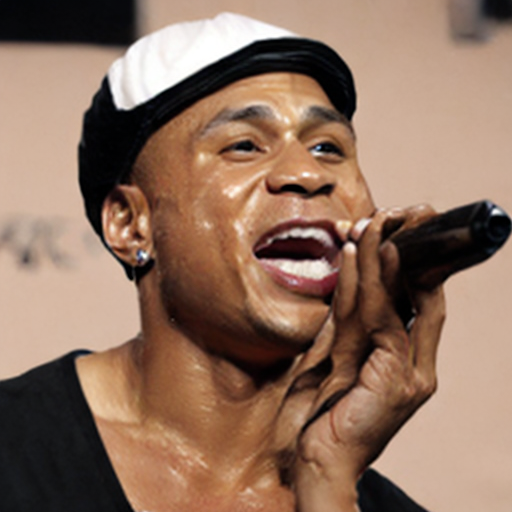}
        \centering
        \end{minipage}
        \begin{minipage}[t]{0.135\linewidth}
        \includegraphics[width=2.44cm]{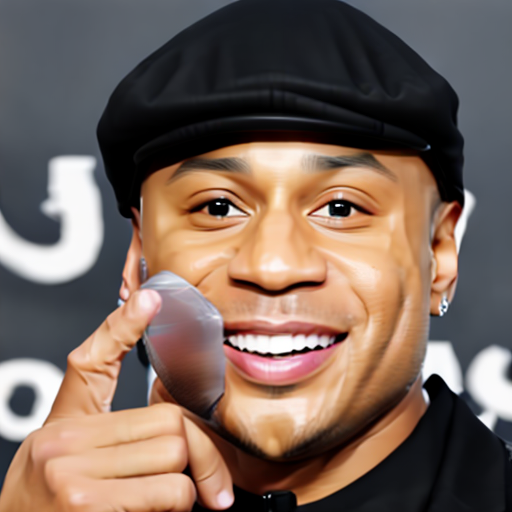}
        \centering
        \end{minipage}
        \begin{minipage}[t]{0.135\linewidth}
        \includegraphics[width=2.44cm]{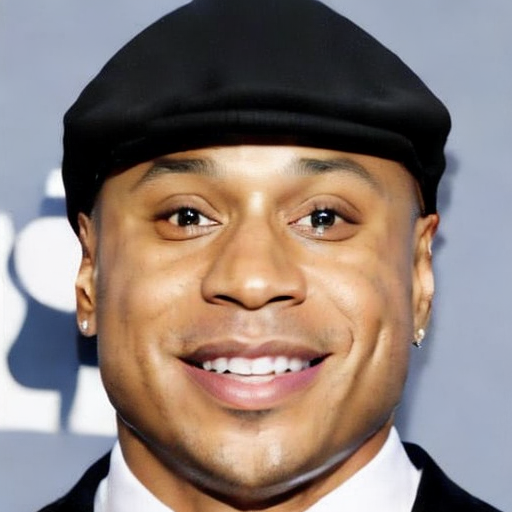}
        \centering
        \end{minipage}
        \end{subfigure}

    \begin{subfigure}{1\textwidth}
    \quad
        \begin{minipage}[t]{0.135\linewidth}
        \centering
        \includegraphics[width=2.44cm]{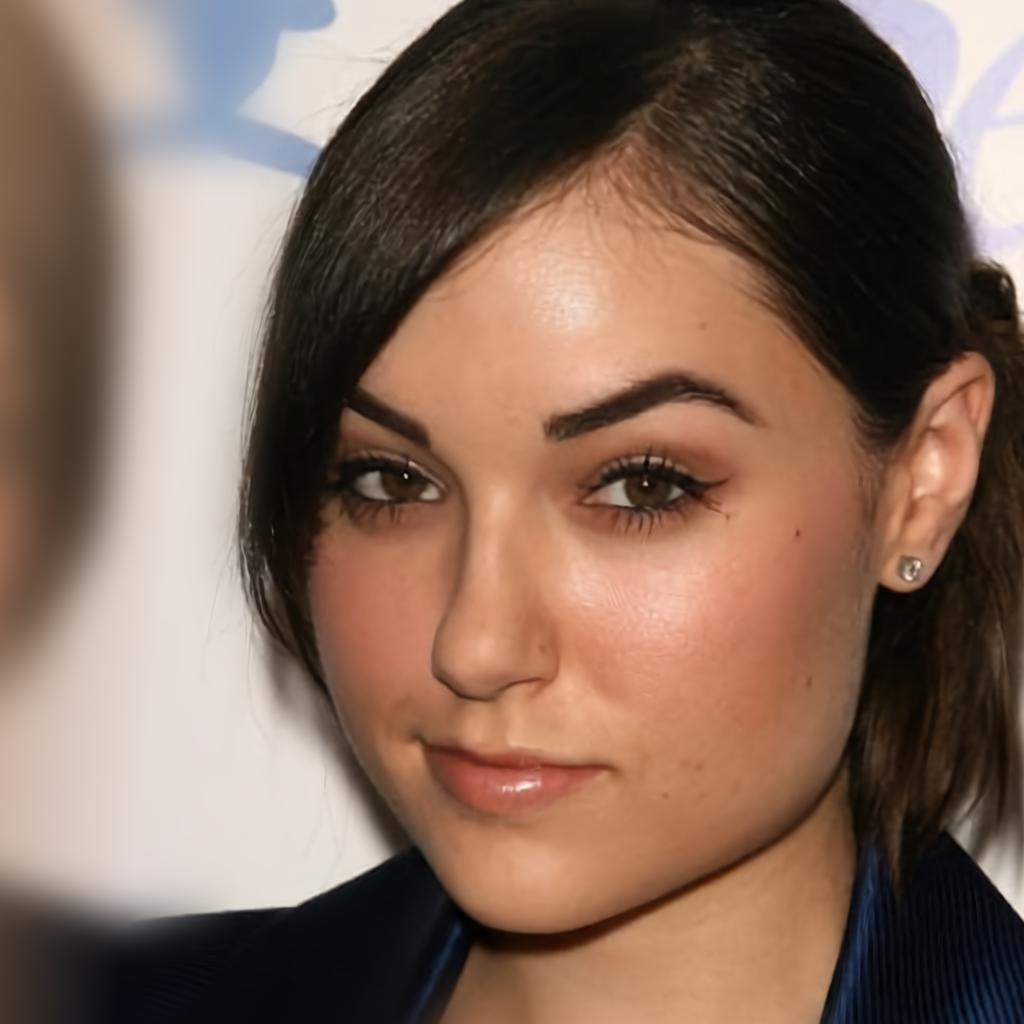}
        \centering
        \end{minipage}
        \begin{minipage}[t]{0.135\linewidth}
        \centering
        \includegraphics[width=2.44cm]{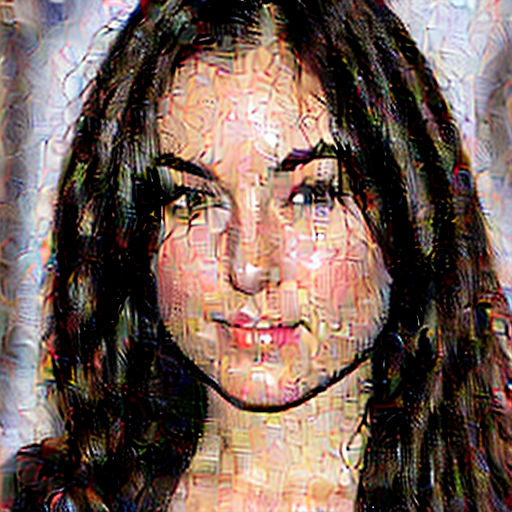}
        \centering
        \end{minipage}
        \begin{minipage}[t]{0.135\linewidth}
        \centering
        \includegraphics[width=2.44cm]{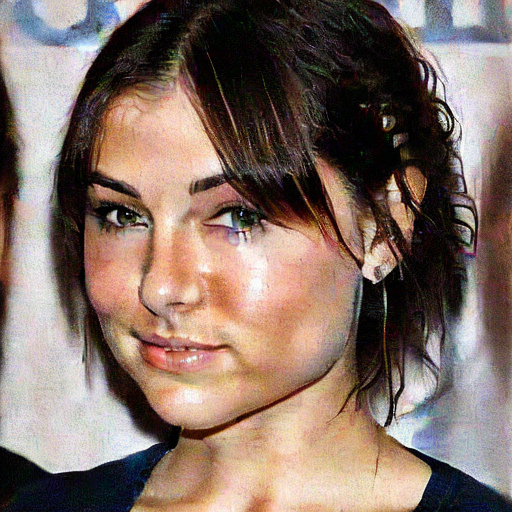}
        \centering
        \end{minipage}
        \begin{minipage}[t]{0.135\linewidth}
        \centering
        \includegraphics[width=2.44cm]{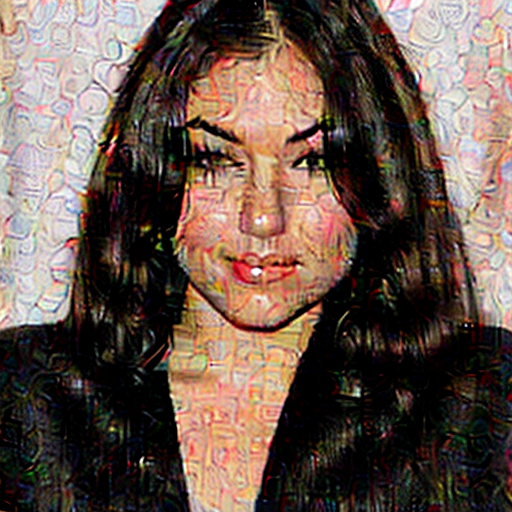}
        \centering
        \end{minipage}
        \begin{minipage}[t]{0.135\linewidth}
        \centering
        \includegraphics[width=2.44cm]{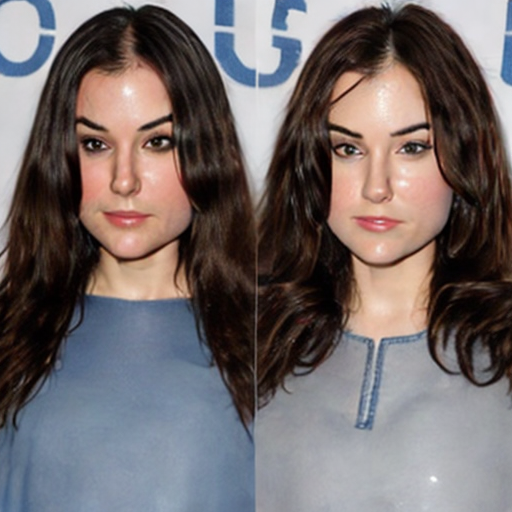}
        \centering
        \end{minipage}
        \begin{minipage}[t]{0.135\linewidth}
        \includegraphics[width=2.44cm]{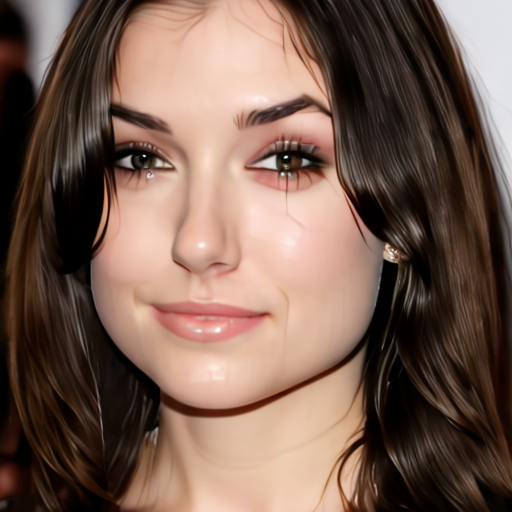}
        \centering
        \end{minipage}
        \begin{minipage}[t]{0.135\linewidth}
        \includegraphics[width=2.44cm]{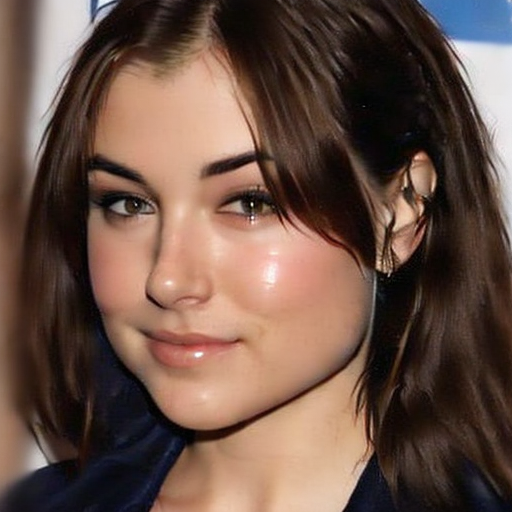}
        \centering
        \end{minipage}
        \end{subfigure}
    
    \begin{subfigure}{1\textwidth}
    \quad
        \begin{minipage}[t]{0.135\linewidth}
        \centering
        \includegraphics[width=2.44cm]{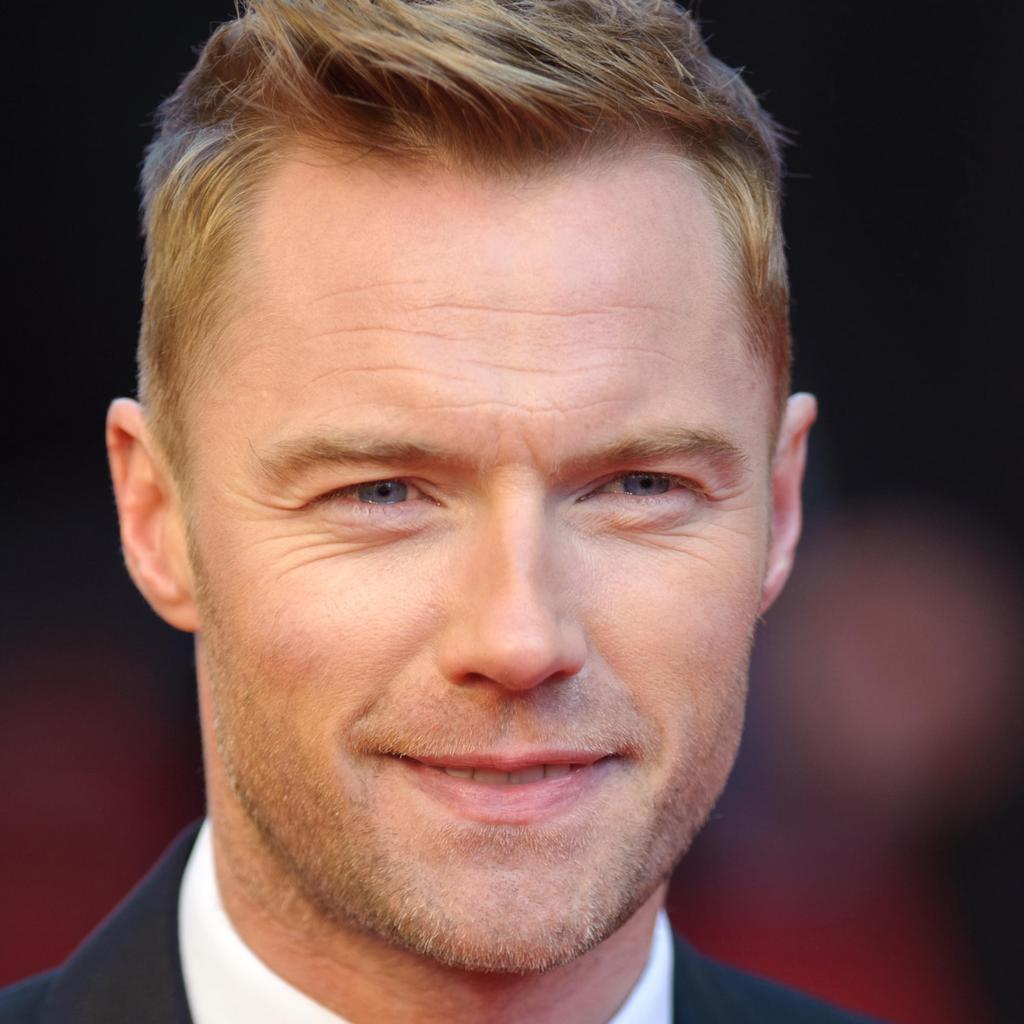}
        \centering
        \caption*{\textbf{Original}}
        \end{minipage}
        \begin{minipage}[t]{0.135\linewidth}
        \centering
        \includegraphics[width=2.44cm]{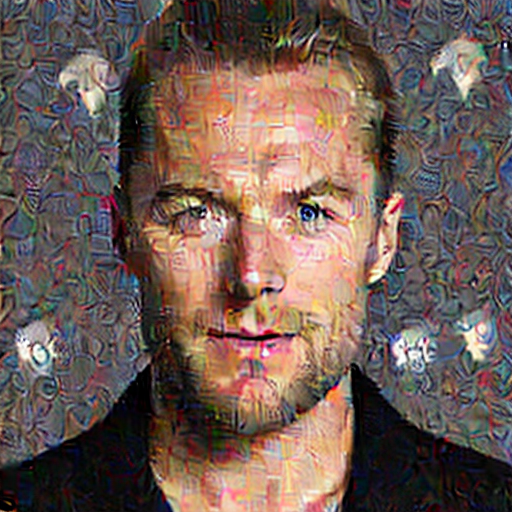}
        \centering
        \caption*{\textbf{w/ def}}
        \end{minipage}
        \begin{minipage}[t]{0.135\linewidth}
        \centering
        \includegraphics[width=2.44cm]{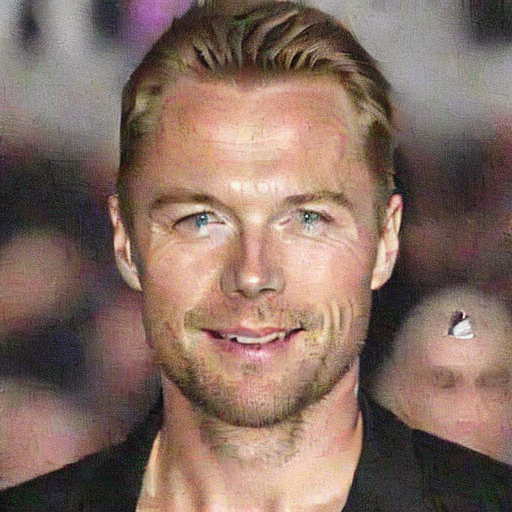}
        \centering
        \caption*{\textbf{Noise}}
        \end{minipage}
        \begin{minipage}[t]{0.135\linewidth}
        \centering
        \includegraphics[width=2.44cm]{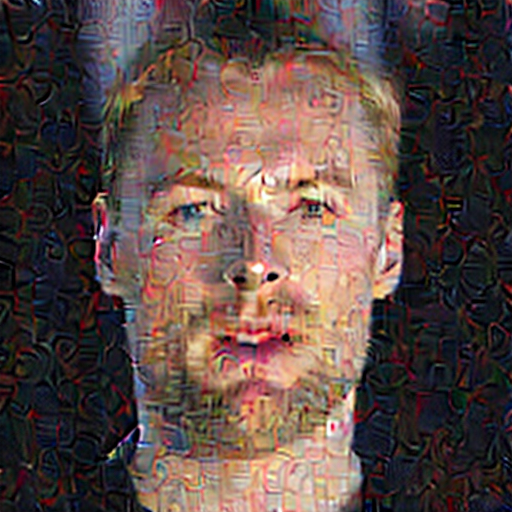}
        \centering
        \caption*{\textbf{IMPRESS}}
        \end{minipage}
        \begin{minipage}[t]{0.135\linewidth}
        \centering
        \includegraphics[width=2.44cm]{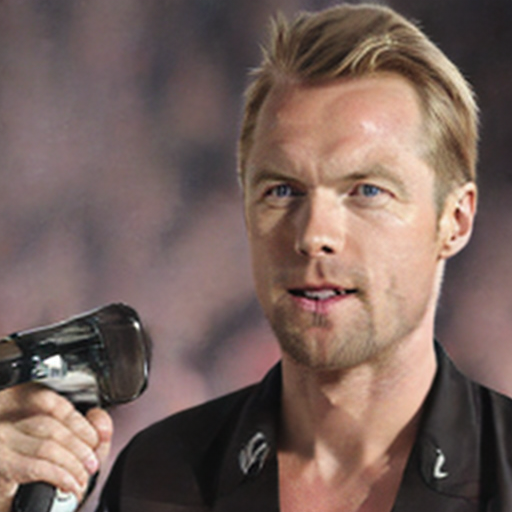}
        \centering
        \caption*{\textbf{DiffPure}}
        \end{minipage}
        \begin{minipage}[t]{0.135\linewidth}
        \includegraphics[width=2.44cm]{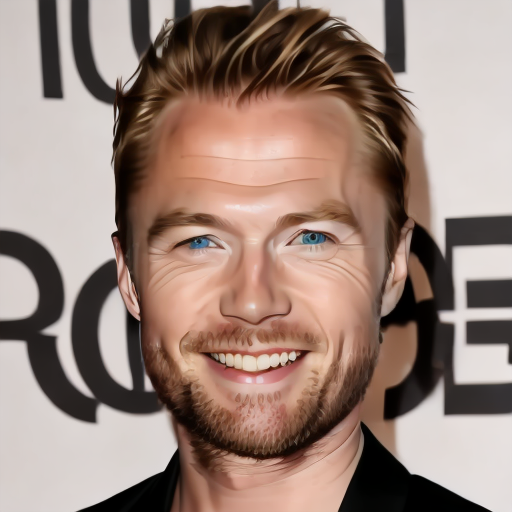}
        \centering
        \caption*{\textbf{GrIDPure}}
        \end{minipage}
        \begin{minipage}[t]{0.135\linewidth}
        \includegraphics[width=2.44cm]{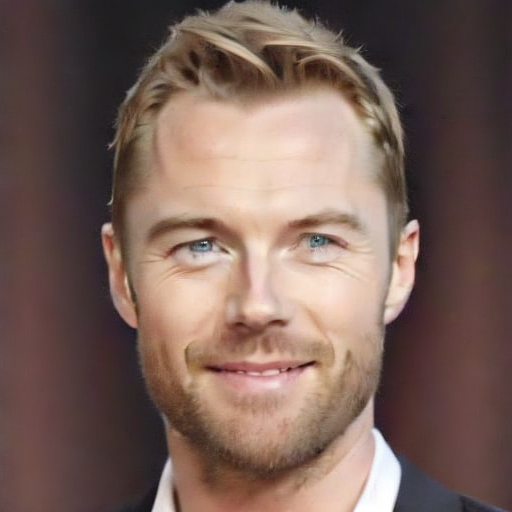}
        \centering
        \caption*{\textbf{TS-LFO}}
        \end{minipage}
        \end{subfigure}

    \caption{Visualized comparison of images generated by different copyright attacks with the original images under the copyright protection of Anti-Dreambooth and SimAC.}
    \label{fig:visualization}
\end{figure*}

\begin{table*}
    \caption{The proportion of selected samples by users on Dreambooth.}
    \label{tab:user_study}
    \centering
    \adjustbox{}{\begin{tabular}{cccccccc}
        \toprule
        \tabcolsep=0.2pt
        Defense & w/def & Noise & IMPRESS & DiffPure & GrIDPure & \gc{\textbf{TS-LFO}} \\
        \midrule
        Anti-Dreambooth & 8.8\% & 49.4\% & 18.1\% & 43.8\% & 40.6\% & \gc{\textbf{71.9\%}} \\
        SimAC & 15.0\% & 36.9\% & 16.3\% & 28.8\% & 63.8\% & \gc{\textbf{78.1\%}} \\
        \bottomrule
    \end{tabular}}
\end{table*}

\subsection{Attack Experiments on More Datasets}
\label{sec:attack_experiments}

% \begin{table*}[t]
%     \caption{The FID performance of our method compared to the performance of other adversarial perturbation removal methods, which may also serve as alternatives for attacking.}
%     \label{tab:more_comparisons}
%     \centering
%     \tabcolsep=2pt
%     \adjustbox{max width=\textwidth}{\begin{tabular}{ccccccccc}
%     \toprule
%     \multirow{2}[0]{*}{Defense} & \multirow{2}[0]{*}{Dataset} & \multirow{2}[0]{*}{w/o def} & \multirow{2}[0]{*}{w/ def} & \multicolumn{5}{c}{Attack} \\
%     \cmidrule(lr){5-9}
%     & & & & JPEG & TVM & MimicDiffusion & IWMF-Diff & \gc{\textbf{TS-LFO}} \\
%     \midrule
%     \multirow{2}[0]{*}{AdvDM} & LSUN-cat & 256.75 & 404.62 & 307.01 & 317.14 & 305.44 & 367.55 &  \gc{\textbf{284.33}} \\
%     & CelebA-HQ & 187.86 & 256.91 & 263.90 & 234.50 & 260.32 & 283.22 &  \gc{\textbf{224.50}} \\
%     \cmidrule(lr){1-9}
%     \multirow{2}[0]{*}{SimAC} & LSUN-cat & 256.75 & 329.38 & 272.30 & 311.71 & 281.92 & 371.78 &  \gc{\textbf{269.47}} \\
%     & CelebA-HQ & 187.86 & 286.98 & 210.94 & 237.85 & 228.66 & 291.82 &  \gc{\textbf{204.80}} \\
%     \cmidrule(lr){1-9}
%     \multirow{2}[0]{*}{DisDiff} & LSUN-cat & 256.75 & 303.13 & 273.31 & 317.55 & 285.70 & 389.67 &  \gc{\textbf{267.34}} \\
%     & CelebA-HQ & 187.86 & 205.94 & 201.08 & 222.55 & 194.58 & 281.43 &  \gc{\textbf{192.67}} \\
%     \bottomrule
%     \end{tabular}}
% \end{table*}

To comprehensively evaluate the attack effectiveness of our method, we design a cross-defense and cross-domain experimental framework. This experimental setup contains two comparative dimensions. In the defense strategy dimension, we select four SOTA defense paradigms: Anti-DreamBooth, AdvDM, SimAC, and DisDiff. In the attack method dimension, we compare our approach against four baseline categories: SOTA adversial purification DiffPure, SOTA copyright attack methods including IMPRESS and GrIDPure, and traditional defense Gaussian noise.

Experimental results in Table \ref{tab:more_comparisons} reveal two critical findings. First, our method demonstrates superior attack performance. On CelebA-HQ against SimAC defense, our method achieves an FID score of 170.15, outperforming DiffPure and GrIDPure by margins of 35.35 and 31.35, respectively. On LSUN-cat, our FID value of 247.57 significantly surpasses the suboptimal DiffPure with a 3.8 improvement.

Furthermore, the traditional Gaussian noise method shows unexpected performance. On the CelebA-HQ dataset, this method can even surpass the SOTA copyright attack methods to become the second-best method behind our TS-LFO. This indicates that there is still great potential for improvement in the performance of current copyright attack and protection methods.

\subsection{Visualization Results and User Study}
\label{sec:visual}

\textbf{Visualization Results.} We provide the visualized comparison of different methods in Fig. \ref{fig:visualization}. From Fig. \ref{fig:visualization}, we observe that our method has successfully learned original image features under all the protection schemes. Compared with the other copyright attack methods, TS-LFO can generate images with significantly higher similarity to the original ones.

\textbf{User Study.} We also conduct the user study to further evaluate the image quality. Specifically, we recruited 100 users and invited them to rate the generated images of different methods. Each user was provided with the following textual instruction: \texttt{``Which images are of the best quality and most similar to the reference (clean) image? Please select 2 to 4 images.''} Then, we analyzed the proportion of samples selected for each method in Table \ref{tab:user_study}. We notice that our TS-LFO has again surpassed all the other methods in the user study.

\subsection{Experiments about Semantic-Level Copyright Protection}
\label{sec:diagnosis_experiments}

In previous experiments, we only evaluated the performance of TS-LFO in pixel-level adversarial noise-based copyright protections. Some methods such as DIAGNOSIS \cite{wangdiagnosis} embed robust watermark signals, generally through semantic-level changes, which may not be attacked. To test the effectiveness of our method against the DIAGNOSIS protection, we conducted the following experiments:

\begin{itemize}
\item[$\bullet$] First, we randomly selected 50 groups of individuals from the CelebA-HQ dataset, with 5 images per group, totally 250 images. These 250 images were protected using the DIAGNOSIS method, resulting in 250 protected images. Using the 250 original images and 250 protected images, we trained a binary classifier for 200 epochs (400 images for the training dataset and 100 images for the validation dataset).

\item[$\bullet$] We then tested the performance of this classifier, and the results are shown in Table \ref{tab:diagnosis_A}. On the 500 images originally used for training (training and validation datasets), the accuracy reached 97.20\%. To further test the classifier's performance, we randomly selected another 50 groups of individuals from CelebA-HQ, obtaining 500 test images in a similar way (test set), and the classifier achieved an accuracy of 92.40\% on these images. This sufficiently demonstrates the superior performance of the trained classifier on the DIAGNOSIS.

\begin{table}
    \caption{The accuracy of the trained classifier in determining whether an image contains a DIAGNOSIS watermark.}
    \label{tab:diagnosis_A}
    \centering
    \adjustbox{max width=\textwidth}{\begin{tabular}{@{}lcccc@{}}
        \toprule
        Metric & train+val & test \\
        \midrule
        prec $\uparrow$ & 97.20\% & 92.40\% \\
        \bottomrule
    \end{tabular}}
\end{table}

\item[$\bullet$] Next, we used the TS-LFO method to perform a copyright attack on the 250 protected images, obtaining 250 attacked images. The 250 original images were divided into 50 groups (5 images per group), and we used DreamBooth to fine-tune a diffusion model, generating 50 images per group, totally 2,500 generated images. These generated images were fed into the classifier to test what percentage of them were classified as containing the DIAGNOSIS semantic watermark. We performed a similar process for the 250 protected images and the 250 attacked images. The experimental results are shown in Table \ref{tab:diagnosis_B}.

\begin{table}
    \caption{The percentage of images flagged with the DIAGNOSIS watermark by the classifier in Table \ref{tab:diagnosis_A}.}
    \label{tab:diagnosis_B}
    \centering
    \adjustbox{max width=\textwidth}{\begin{tabular}{@{}lcccc@{}}
        \toprule
        Metric & original & protected & attacked \\
        \midrule
        percentage $\uparrow$ & 2.32\% & 17.24\% & 0.92\% \\
        \bottomrule
    \end{tabular}}
\end{table}

\item[$\bullet$] For the images generated from the original images, only 2.32\% were classified by the classifier as containing the DIAGNOSIS semantic watermark. In contrast, for the images generated from the protected images, 17.24\% were classified as containing the DIAGNOSIS watermark. This indicates that the DIAGNOSIS defense significantly increases the probability of the generator producing images with the semantic watermark embedded. (Note: In the DIAGNOSIS paper, this ratio was 100\% rather than 17.24\%. This discrepancy arises because the DIAGNOSIS authors only experimented with LoRA and LoRA+DreamBooth generative models, not the standalone DreamBooth model we used here. Additionally, DIAGNOSIS typically fine-tunes diffusion models using hundreds to tens of thousands of images, whereas we used only 5 images per group to fine-tune the model. Fine-tuning with just 5 images was insufficient for the diffusion model to fully learn the characteristics of the semantic watermark.)

\item[$\bullet$] After applying TS-LFO, this proportion decreased to 0.92\%, even lower than the baseline from the original images. This demonstrates that TS-LFO can effectively attack the DIAGNOSIS copyright defense method.

\end{itemize}

\newpage

\begin{algorithm}[h]
\caption{TS-LFO Variant with Gaussian Latent Distribution}
\label{alg:ts_lfo_gaussian}
\begin{algorithmic}[1]
\Require
    \State $N$: Number of training steps
    \State $x_{adv}$: Adversarially perturbed input image
    \State $E(\cdot)$: Encoder outputting Gaussian parameters $(\mu, \sigma)$
    \State $D(\cdot)$: Pre-trained LDM decoder
    \State $\epsilon_{\theta}(\cdot)$: Pre-trained Unet denoiser
    \State $M(\cdot,\cdot)$: An image similarity metric
    \State $T$: Total diffusion timesteps
    \State $\eta$: Learning rate
    \State $\lambda_l, \lambda_r$: Initial/Final weight factors
    \State $\epsilon_{rec}$: Reconstruction threshold
    
\Ensure
    \State $z_{new}$: Optimized latent feature
    
\Procedure{TS-LFO-Gaussian}{$N, x_{adv}, E, D, \epsilon_{\theta}, M, T, \eta, \lambda_l, \lambda_r, \epsilon_{rec}$}
    
    \State \textcolor{gray}{\textit{// Stage 1: Latent Denoising }}
    \State \textcolor{gray}{\textit{// Initial distribution parameters }}
    \State $(\mu, \sigma) \gets E(x_{adv})$  
    
    \For{$i = 1$ \textbf{to} $N$}
        \State Sample $t$ from $[1,T]$ randomly
        \State \textcolor{gray}{\textit{// Reparameterization}}
        \State Sample $z \sim \mathcal{N}(\mu, \sigma^2)$
        \State \textcolor{gray}{\textit{// Compute $\alpha_t$ via noise schedule}}
        \State $z_t \gets \sqrt{\alpha_t}z + \sqrt{1-\alpha_t}\epsilon$
        
        \State \textcolor{gray}{\textit{// Loss Computation}}
        \State $\mathcal{L}_{align} \gets M(D(z),x_{adv})$ 
        \State $\mathcal{L}_{DM} \gets \|\epsilon - \epsilon_{\theta}(z_t, t, c)\|_2^2$
        \State $\lambda_t \gets \lambda_l + \frac{(\lambda_r - \lambda_l)}{T} \cdot t$
        \State $\mathcal{L}_{total} \gets \mathcal{L}_{align} + \lambda_t\mathcal{L}_{DM}$
        
        \State \textcolor{gray}{\textit{// Distribution Update}}
        \State $\nabla_{\mu} \gets \frac{\partial \mathcal{L}_{total}}{\partial \mu}$ 
        \State $\nabla_{\sigma} \gets \frac{\partial \mathcal{L}_{total}}{\partial \sigma}$
        \State \textcolor{gray}{\textit{// Update mean}}
        \State $\mu \gets \mu - \eta \nabla_{\mu}$ 
        \State \textcolor{gray}{\textit{// Update std dev}}
        \State $\sigma \gets \sigma - \eta \nabla_{\sigma}$ 
    \EndFor
    
    \State \textcolor{gray}{\textit{// Stage 2: Latent Reconstruction}}
    \State \textcolor{gray}{\textit{// Final sampling}}
    \State Sample $z \sim \mathcal{N}(\mu, \sigma^2)$  
    \State $x_{rec} \gets \mathrm{clamp}(D(z), x_{adv}-\epsilon_{rec}, x_{adv}+\epsilon_{rec})$
    \State \textcolor{gray}{\textit{// Re-encode with new distribution}}
    \State $z_{new} \gets E(x_{rec}).\mathrm{sample}()$ 
    
    \State \Return $z_{new}$
\EndProcedure
\end{algorithmic}
\end{algorithm}